\documentclass[%
twocolumn,
superscriptaddress,
nofootinbib,
amsmath,amssymb,
aps,
prd,
]{revtex4-1}

\usepackage{graphicx}
\usepackage{float}% to force figure placement 
\usepackage{bm}% bold math
\usepackage[hidelinks]{hyperref}
\usepackage{xcolor}

\usepackage[utf8]{inputenc}
\usepackage[T1]{fontenc}
\usepackage{slashed}

\usepackage{fullpage}
\usepackage{enumerate}

\usepackage{mathalfa,amssymb,amsthm}
\usepackage{amsmath}

\begin{document}

%\preprint{APS/123-QED}

\title{Black hole-neutron star mergers 
in Einstein-scalar-Gauss-Bonnet gravity}
\author{Maxence Corman}
\email{maxence.corman@aei.mpg.de}
\affiliation{Max Planck Institute for Gravitational Physics (Albert Einstein Institute), D-14476 Potsdam, Germany}
\author{William E. East}
\email{weast@perimeterinstitute.ca}
\affiliation{%
Perimeter Institute for Theoretical Physics, Waterloo, Ontario N2L 2Y5, Canada.
}%

%-----------------------------------------------------------------------------

\date{\today}% It is always \today, today,

\begin{abstract}
    Gravitational wave observations of black hole-neutron star binaries,
    particularly those where the black hole has a lower mass compared to other
    observed systems, have the potential to place strong constraints on
    modifications to general relativity that arise at small curvature length
    scales. Here we study the dynamics of black hole-neutron star mergers in
    shift-symmetric Einstein-scalar-Gauss-Bonnet gravity, a representative
    example of such a theory, by numerically evolving the full equations of
    motion. We consider quasi-circular binaries with different mass-ratios
    that are consistent with recent gravitational wave observations,
    including cases with and without tidal disruption of the star, and quantify
    the impact of varying the coupling controlling deviations from general
    relativity on the gravitational wave signal and scalar radiation.  We find
    that the main effect on the late inspiral is the accelerated frequency
    evolution compared to general relativity, and that---even considering Gauss-Bonnet
    coupling values approaching those where the theory breaks down---the impact
    on the merger gravitational wave signal is  mild, predominately 
    manifesting as a
    small change in the amplitude of the ringdown. We compare our results to
    current post-Newtonian calculations and find consistency 
    throughout the inspiral.
\end{abstract}

\maketitle

%=============================================================================
%\allowdisplaybreaks
%\tableofcontents
%=============================================================================
\section{Introduction}
\label{sec:introduction}
The growing number of gravitational wave observations \cite{LIGOScientific:2016dsl,
LIGOScientific:2018mvr,LIGOScientific:2020ibl,LIGOScientific:2021djp} 
has opened up new opportunities to
probe the strong gravity regime governing the coalescence of compact 
binary systems, marking a new chapter for tests of General Relativity (GR)
\cite{Will:2014kxa,Yunes:2013dva,Yunes:2016jcc,Berti:2015itd,Abbott:2018lct,
LIGOScientific:2019fpa,LIGOScientific:2020tif,LIGOScientific:2021sio,Berti:2018cxi,
Berti:2018vdi,TheLIGOScientific:2016src}. 
To robustly test GR in this highly dynamical and nonlinear regime,
we need accurate predictions in modified theories of gravity that cover
the full inspiral, merger, and ringdown of compact object binaries.

Many proposed modifications to GR rely on effective field theory arguments 
\cite{Donoghue:1994dn,Burgess:2003jk}, and hence result from adding additional
curvature terms to the Einstein-Hilbert action multiplied by some coupling constants
with (positive) dimensions of length. Examples include
not only the most generic Horndeski theories \cite{Horndeski:1974wa} and 
dynamical Chern-Simons gravity \cite{Alexander:2009tp},
but also theories that add higher-dimensional curvature operators 
without introducing new light degrees of freedom \cite{Endlich:2017tqa,deRham:2020ejn}.
It is then natural to expect that such alternative theories of gravity 
exhibit the strongest
deviations in the presence of the shortest curvature length scales. This makes the 
smallest mass compact objects ideal probes for finding evidence of, or constraining
such theories.

In this paper, we study how black hole-neutron star (BHNS) mergers can be used
to probe a representative modified theory of gravity introducing modifications to GR
at small curvature length scales, namely Einstein-scalar-Gauss-Bonnet (EsGB) gravity.
An interesting aspect of EsGB gravity is that neutron stars carry no scalar charge,
while black holes do 
\cite{Kanti:1997br,Yunes:2011we,Sotiriou:2013qea,Sotiriou:2014pfa,Antoniou:2017acq,
Papageorgiou:2022umj}. 
This means that unlike
Damour-Esposito-Farese scalar-tensor (ST) theories \cite{Damour:1993hw},
where neutron stars develop a scalar charge and black holes do not
\cite{Damour:1992we,Anderson:2019hio,Hawking:1972qk,Sotiriou:2011dz},
EsGB gravity evades binary pulsar system constraints based on dipolar
radiation~\cite{Damour:1996ke,Will:2014kxa}%
\footnote{We note that while ST theories are strongly constrained
by binary pulsar observations \cite{Damour:1996ke,Will:2014kxa}, there are examples where 
the neutron star undergoes \emph{spontaneous scalarization} (dynamical \cite{Barausse:2012da,Palenzuela:2013hsa} or 
induced \cite{Damour:1992we,Damour:1993hw}) or the scalar field is massive~\cite{Ramazanoglu:2016kul},
which suppresses effects at the separations currently observed, hence avoiding current constraints.}%
, and instead one has to search
for observational signatures in other ways, such as through 
compact object merger dynamics.
An important feature of EsGB gravity
is the emission of scalar radiation in addition
to the usual tensor radiation found in GR \cite{Yagi:2011xp}. 
Similar to ST theories, the leading scalar radiation is dipolar, and 
thus dominates over quadrupolar gravitational 
waves at low frequencies \cite{Sennett:2016klh}. The strength
of this radiation depends not only on the scalar charge of the compact objects,
which is inversely proportional to the square of the smallest mass black hole 
in the system,
but also on the square of the difference between the charges of the
constituent objects~\cite{Sennett:2016klh}. 
Therefore, if two objects possess similar charges, such as 
in an equal or near equal mass binary black hole merger, the dipolar radiation will be 
suppressed. Conversely, the strongest constraints on EsGB gravity, as well as
ST theories, can come from a mixed
binary consisting of ideally a small black hole and a neutron star, as only one of them
carries a scalar charge.
Recently, in Ref.~\cite{Ma:2023sok}, fully nonlinear numerical simulations were used to study a BHNS merger in ST theory\footnote{ST theories do not modify the principal part
of the Einstein equations and can be evolved in the same way.},
focusing on
how the gravitational wave emission is affected by the spontaneous scalarization 
of the neutron star. 
Considering a binary with parameters consistent with the gravitational wave event GW200115
\cite{LIGOScientific:2021qlt}, Ref.~\cite{Ma:2023sok} found that the ST system inspiraled faster than its GR
counterpart due to the emission of scalar radiation, showing good agreement with predictions from 
post-Newtonian (PN) theory during the inspiral. 

In the third observing run of the Advanced LIGO \cite{LIGOScientific:2014pky}, Advanced
Virgo \cite{VIRGO:2014yos}, and KAGRA \cite{Somiya:2011np,Aso:2013eba} network of
gravitational wave
detectors,
the LIGO-Virgo detectors observed GW200105 and GW200115,
the first gravitational wave detections from the mergers of BHNS systems
\cite{LIGOScientific:2021qlt}. By adopting the leading order (dipolar) PN
correction, 
Ref.~\cite{Lyu:2022gdr} derived the strongest bound on coupling constant 
of theory at that time to be on the order of
kilometer or less. This was achieved
through a Bayesian Markov-chain Monte Carlo analysis 
combining GW200105, GW200115, GW190814~\cite{LIGOScientific:2020zkf}
\footnote{This event is consistent
with both a binary black hole and a BHNS binary.}, and selected binary
black hole events.
More recently, the LIGO-Virgo-KAGRA Collaboration (LVK) reported 
the observation of a compact object binary merger in May 2023, GW230529, with component masses
$3.6^{+0.8}_{-1.2} M_{\odot}$ and $1.4^{+0.6}_{-0.2} M_{\odot}$, the most probable
interpretation of which is the coalescence of a black hole in the lower mass gap and
a neutron star \cite{LIGOScientific:2024elc}.
To verify whether GW230529 is consistent with GR, the LVK collaboration performed
so-called parameterized tests, searching for parametric deviations to the gravitational
wave phase during the inspiral, specifically using the TIGER 
\cite{Agathos:2013upa,Meidam:2017dgf} and FTI 
\cite{Mehta:2022pcn} frameworks. For all waveform models and PN orders so far considered,
GW230529 was found to be consistent with GR, with constraints on the -1PN 
(dipolar) coefficient being an order of magnitude tighter than previous
bounds for BHNS and binary black hole (BBH) mergers reported by the collaboration 
\cite{LIGOScientific:2024elc}. Applying the same methods as Ref.~\cite{Lyu:2022gdr}
to GW230529, Ref.~\cite{Gao:2024rel} improved the bounds on coupling constant by 
a factor of approximately four. Alternatively,
mapping the -1PN constraints from FTI tests to a constraint on the
coupling constant in EsGB gravity, Ref.~\cite{Sanger:2024} also 
obtained a tighter bound on
coupling constant.
With upcoming improvements in the 
sensitivity of the LVK gravitational wave detectors \cite{KAGRA:2013rdx},
as well as future third-generation ground-based detectors 
\cite{Hild:2010id,LIGOScientific:2016wof},
it is thus timely and vital to provide predictions of 
gravitational wave signals from BHNS binaries in EsGB gravity.
Although significant progress has been made in modeling
compact object mergers in EsGB gravity
using PN theory \cite{Yagi:2011xp,Sennett:2016klh,Julie:2019sab,Shiralilou:2021mfl,Shiralilou:2020gah}, 
as one approaches the merger, PN theory breaks down, and numerical relativity is required.
For EsGB gravity, numerical relativity has been used to study 
 binary neutron star and 
binary black hole systems, solving the full equations~\cite{East:2022rqi,East:2020hgw,Ripley:2022cdh,Corman:2022xqg,AresteSalo:2022hua,AresteSalo:2023mmd},
and using a decoupling or order-by-order approximation~\cite{Witek:2018dmd,Silva:2020omi,Okounkova:2020rqw,Kuan:2023trn}.
See also Ref.~\cite{Corman:2024cdr} for a comparison of the different approaches 
for treating modifications to full general relativity that have been used to study binary black hole mergers.

However, a simulation of a BHNS merger in EsGB gravity is still missing. 
In this work, we aim to fill this gap
and take advantage of recent advances in solving the full equations of 
shift-symmetric EsGB gravity to
study the nonlinear dynamics of BHNS mergers in this theory. 
In particular, we make use of the modified harmonic
formulation \cite{Kovacs:2020pns,Kovacs:2020ywu} 
and methods developed in Refs.~\cite{East:2020hgw,East:2022rqi,Corman:2022xqg} 
for evolving black holes and neutron stars in shift-symmetric EsGB gravity.
We focus on how the gravitational wave
emission is impacted by the presence of an additional energy dissipation channel.

Motivated by the LVK observations, we consider two binary systems, namely one consistent with
GW200115, and another with GW230529. For the latter, we choose an equation of state (EOS) and mass ratio
so that the neutron star is tidally disrupted. 
For coupling values comparable to the upper bound obtained from GW200115, 
we find a noticeable dephasing in the gravitational
wave signal compared to GR in the inspiral. However, in the last few orbits, the rate of dephasing becomes
small (even being consistent with zero or indicating a slower inspiral rate for EsGB compared to GR to within the
numerical errors). In part due to this suppression, we find the PN approximation to 
be consistent with our results into the late inspiral.
We also study the effect of modifications to GR on the merger and ringdown signal.
We find that the effect on the peak amplitude of the gravitational
wave signal is small, with the relative 
change in the ringdown frequency being at most on the order of 
$\sim 1\%$ for the largest couplings
we consider, while the amplitude of ringdown signal can vary 
by $\sim 10\%$ for the GW200115-like binary.
We observe a amplification (or suppression) in the amplitude of ringdown
gravitational wave signal with increasing coupling when neutron star is (or not) tidally disrupted.
We conjecture that the amplification is due to the neutron star being more compact
and less strongly tidally disrupted for larger couplings.
Finally, in the case where neutron star is tidally disrupted, we find that the amount of mass remaining
outside black hole after the merger decreases slightly with increasing coupling when fixing the EOS. 

The remainder of this paper is organized as follows. 
We review the theory we consider, shift-symmetric EsGB gravity,
in Sec.~\ref{sec:eom_derivation}. 
We describe our numerical methods for evolving this theory coupled to hydrodynamics
and analyzing the results in Sec.~\ref{sec:numerical_methods}. 
Results from our study of BHNS binaries in shift-symmetric
EsGB gravity are presented in Sec.~\ref{sec:results}. 
We discuss these results and conclude in Sec.~\ref{sec:conclusion}.
We discuss the accuracy our simulations in Sec.~\ref{app:convergence}.
We use geometric units:
$G=c=1$, a metric sign convention of $-+++$, and lower case Latin letters to index
spacetime indices.  
The Riemann tensor is $R^a{}_{bcd}=\partial_c\Gamma^a_{db}-\cdots$.

%=============================================================================
%=============================================================================
\section{Shift-symmetric Einstein Scalar Gauss Bonnet gravity}
\label{sec:eom_derivation}
The action for shift-symmetric EsGB gravity is:
\begin{align}
\label{eq:leading_order_in_derivatives_action}
    S
    =&
    \frac{1}{16\pi}\int d^4x\sqrt{-g}
   \left(
        R
    -   \left(\nabla\phi\right)^2
    +	2\lambda\phi\mathcal{G}
	\right) + S_{\mathrm{matter}}
    ,
\end{align}
where $g$ is the determinant of spacetime metric and $\mathcal{G}$ is the Gauss-Bonnet scalar:
\begin{align}
    \mathcal{G}
    \equiv
    R^2 - 4R_{ab}R^{ab} + R_{abcd}R^{abcd}
    .
\end{align}
Here, $\lambda$ is a constant coupling parameter that, in
geometric units, has dimensions of length squared, $\phi$ is the scalar field 
and $S_{\rm{matter}}$ is
the action for any other matter (in our case the neutron star fluid).
As the Gauss-Bonnet scalar $\mathcal{G}$ is a total derivative
in four dimensions, we see that the action 
is preserved
up to total derivatives under constant shifts in the scalar field:
$\phi\to\phi+\textrm{constant}$.

The covariant equations of motion for shift-symmetric EsGB gravity are
\begin{align}
\label{eq:eom_esgb_scalar}
   \Box\phi
   +  
   \lambda\mathcal{G}
   &=
   0
   ,\\
\label{eq:eom_edgb_tensor}
   R_{ab}
   -  
   \frac{1}{2}g_{ab}R
   + 2\lambda
   \delta^{efcd}_{ijg(a}g_{b)d}R^{ij}{}_{ef}
   \nabla^g\nabla_c\phi
   &=
   8 \pi T_{ab}
   ,
\end{align}
where $\delta^{abcd}_{efgh}$ is the generalized Kronecker delta tensor and 
$T_{ab} =  T_{ab}^{\rm SF} + T_{ab}^{\rm matter} $ with
\begin{equation}\label{eq:Tsf}
T_{ab}^{\mathrm{SF}} \equiv \frac{1}{8\pi} \left(  
   \nabla_a\phi\nabla_b\phi -  
   \frac{1}{2}\left(\nabla\phi\right)^2g_{ab}\right) \ .
\end{equation}
We do not introduce any non-minimal coupling for the matter in the Einstein frame,
and the equations of motion for the matter terms are the same as in GR.

Schwarzschild and Kerr black holes are not stationary solutions 
in this theory: if one begins with such vacuum initial data, the black holes will 
dynamically develop stable scalar clouds (hair).
The end state is a black hole with nonzero scalar charge $Q_{\rm SF}$,
such that at large radius the scalar field falls of like
$\phi = Q_{\rm SF}/r + \mathcal{O}(1/r^2)$.
Furthermore, studies have found that stationary solutions exist,
as long as the coupling normalized by the total black hole mass as measured
at infinity $m$,
$\lambda/m^2$, is sufficiently small
\cite{Sotiriou:2013qea,Sotiriou:2014pfa,Ripley:2019aqj,East:2020hgw}.
In particular, 
regularity of black hole solutions and hyperbolicity of the theory set
$\lambda/m^2\lesssim 0.23$ for non-spinning black holes, 
\cite{Sotiriou:2014pfa,Ripley:2019aqj}.

Neutron stars, in contrast to black holes, do not have a scalar charge in EsGB gravity 
\cite{Yagi:2011xp,Julie:2019sab}.
However, it is important to note that despite not having a scalar charge,
a neutron star in EsGB gravity will still be surrounded by a scalar cloud (sourced
by the Gauss-Bonnet invariant). 
The lack of scalar charge arises because the scalar field decays much
more rapidly than with $1/r$, as would be required for the neutron star to have a 
scalar charge. Neutron stars in shift-symmetric EsGB gravity were studied 
(restricting to spherical symmetry) in Ref.~\cite{Pani:2011xm},
where it was numerically found that (independently of the equation of state of neutron star)
turning on the
EsGB coupling tends to reduce the maximum gravitational mass or 
increase the central density 
when a solution exists. It was further analytically 
shown that depending on the value of the coupling and neutron star EOS,
there is a maximum central
density beyond which no spherically symmetric perfect fluid solutions 
can be constructed.

In general, the equations of motion for EsGB gravity can only be evolved in 
time in a well-posed manner for weakly-coupled
solutions \cite{Ripley:2019irj,Kovacs:2020pns,Kovacs:2020ywu,East:2020hgw,Ripley:2022cdh} 
\footnote{Weak coupling means that the Gauss-Bonnet corrections to the equations 
of motion remain
sufficiently small compared to the leading two-derivative terms. 
This is consistent with strong-field
black hole dynamics provided the size of the smallest black hole in the system is larger than the length scale.}.
In earlier studies of collapse and black hole and neutron star binaries in EsGB gravity, 
it was found
that when the coupling of the theory is made too large, the compact objects can evolve 
from an initial weakly coupled state,
to a strongly coupled state, where the hyperbolicity of the evolution equations breaks down,
although approaching this limit does not appear to be
preceded by any singular behavior developing in the metric or scalar field 
\cite{Ripley:2019hxt,Ripley:2019irj,
Ripley:2020vpk,East:2020hgw,East:2022rqi,East:2021bqk,Corman:2022xqg}.
Here we find evidence that this breakdown happens in BHNS mergers not only when a black hole scalarizes, but 
also sometimes when
the scalar field in the neutron star grows in magnitude, leading to an increase in the star's density.
However, this breakdown occurs for 
coupling values comparable to or larger than the best
existing constraints on EsGB gravity and approaching this limit does not appear to 
be preceded by dramatically
different spacetime and/or scalar field dynamics.

%=============================================================================
\section{Methods}
\label{sec:numerical_methods}
%------------------------------------------------------------------------------
\subsection{Evolution equations and code overview}
\label{sec:evolution_methods}

We numerically evolve the full shift-symmetric EsGB equations of motion
using the modified generalized harmonic 
formulation~\cite{Kovacs:2020pns,Kovacs:2020ywu}.
We use similar choices for the gauge and numerical parameters as in
Ref.~\cite{East:2020hgw}.
We model the neutron star using ideal hydrodynamics. The Euler equations are
the same as in GR, and evolved using the hydrodynamics code described in Ref.~\cite{East:2011aa}.
We use the same methods and parameters for evolving BHNS binaries as in
Ref.~\cite{East:2011xa}.
Our simulations use box-in-box adaptive mesh refinement provided by
the PAMR library \cite{PAMR_online}. We typically use seven 
levels of mesh refinement in our simulations, unless otherwise noted.
We provide details on numerical resolution and convergence in
Appendix~\ref{app:convergence}.
%------------------------------------------------------------------------------
\subsection{Initial data and cases considered\label{sec:initial_data}}
\begin{table*}[t]
\centering
\begin{tabular}{ |p{1.7cm}||p{1.5cm} p{1.5cm} p{1.5cm} p{1.5cm} p{1.5cm} p{1.5cm} p{1.5cm} p{2cm}|   }
 \hline
	GW event& $m_{\rm BH}/M_{\odot}$ & $m_{\rm NS}/M_{\odot}$ & $q$ &$R_{\rm NS}/\rm{km}$ & $D/M$ & TD & $N^{\rm GR}_{\rm cycle}$ & $\sqrt{\alpha_{\rm GB}}/{\rm km}$\\
 \hline
	GW200115   & 5.7    &1.5&   0.26&12.3 &10.35 & No &7 & $\{0,1.19\}$\\
	GW200115&  5.7  & 1.5   &0.26& 12.3 & 8.61 & No & 4 &$\{0,0.84,1.19\}$\\
	GW230529 &3.5 & 1.4&  0.40& 12.3& 9.82 &Yes &5  &$\{0,0.73,0.89\}$\\
 \hline
\end{tabular}
	\caption{Summary of the parameters of the GW200115- and GW230529-like
	BHNS systems we consider. The black hole is non-spinning and
	has an irreducible mass
	$m_{\rm BH}$, while the neutron star has a gravitational 
	mass $m_{\rm NS}$ with radius
	given by $R_{\rm NS}$. TD indicates whether neutron star is tidally disrupted
	or not before merger and $N^{\rm GR}_{\rm cycle}$ is the number of gravitational wave 
	cycles before merger in GR. The coupling values we consider are denoted by
	$\sqrt{\alpha_{\rm GB}}$.}
	\label{tab:summary}
\end{table*}

We use quasi-circular BHNS binary initial data constructed with
the Frankfurt University/Kadath (\texttt{FUKA}) Initial Data code suite 
\cite{Papenfort:2021hod}, 
which is based on an extended version of the \texttt{KADATH} spectral solver library 
\cite{Grandclement:2009ju}. We choose $\phi=\partial_t \phi=0$ on the initial time slice,
in which case the constraint equations of shift-symmetric EsGB gravity reduce 
to those of vacuum GR. We slowly ramp on the coupling of the theory as described
in Appendix B of Ref.~\cite{Okounkova:2019zjf}, in such a way that the scalar field grows
on a timescale that is short compared with the orbital binary timescale.
The set of hydrodynamical evolution equations 
are closed by an equation of state connecting pressure $p$
to specific internal energy $\epsilon$ and rest mass density $\rho$, i.e., $p=p(\rho,\epsilon)$.
Though in general it would be interesting to consider different EOSs in order
to determine how this impacts the results and to test for possible degeneracies,
for this first study we consider a single one for the neutron star.
We use a cold piecewise polytropic EOS \cite{Read:2008iy} 
approximating the ALF2 EOS \cite{Alford:2004pf} for the neutron star. 
This prototypical
stiff EOS predicts the radius of a $1.4\ M_{\odot}$ neutron star to be $R_{1.4} \sim 12.32$ km 
\cite{Gonzalez:2022mgo},
has a maximum mass of $\sim 2.0 M_{\odot}$ for non-spinning stars~\cite{Alford:2004pf},
and is consistent with pulsar observations
\cite{NANOGrav:2019jur,Riley:2021pdl,van_Kerkwijk_2011,Fonseca:2016tux,Antoniadis:2013pzd},
with both electromagnetic and gravitational wave 
observations \cite{Margalit:2017dij,Radice:2017lry,LIGOScientific:2018hze,Pang:2021jta,
Coughlin:2018fis,Radice:2018ozg} 
of the binary neutron star event GW170817 \cite{TheLIGOScientific:2017qsa},
as well as the gravitational wave observations of GW190814 \cite{Tan:2020ics} and GW190425 
\cite{Fasano:2020eum}.
Thermal effects are added to the zero-temperature
polytrope with an additional pressure contribution of the form 
$p_{\rm th} = (\Gamma_{\rm th}-1)\rho\epsilon$,
where $\epsilon_{\rm th}$ denotes the excess specific energy compared
to the cold value at the same density. We use $\Gamma_{\rm th} =1.75$,
motivated by studies comparing non-zero temperature EOSs such as Refs.~\cite{Yasin:2018ckc,Bauswein:2010dn}.

The binary parameters we consider for one of the BHNS system we study
are chosen to be consistent
with GW200115 \cite{LIGOScientific:2021qlt}.
The source of GW200115 has component masses $5.7^{+1.8}_{-2.1}$ and
$1.5^{0.7}_{-0.3}\ M_{\odot}$ at a $90\%$ confidence level and mass ratio
$q=0.26^{0.35}_{-0.10}$. The primary spin has a 
negative spin projection onto the orbital angular momentum (anti-aligned spin),
but is also consistent with zero spin $\chi_1=0.33^{+0.48}_{-0.29}$.
The spin and tidal deformability of the neutron star were unconstrained, 
and no electromagnetic
counterpart has been identified to date. We consider a non-rotating neutron star 
with gravitational mass $m_{\rm NS} = 1.5 M_{\odot}$ 
and a non-spinning black hole with mass
$m_{\rm BH}=5.7 M_{\odot}$. We consider two initial separations: 
$D = 10.35M$ and $D = 8.61M$, 
where $M=7.2\ M_{\odot}$ is total mass of system. The systems undergo approximately
7 and 4.5 orbital periods, respectively, before merging in GR. For the longer inspiral,
we consider two values of the coupling parameter, namely $\lambda/m_{\rm BH}^2=(0,0.1)$.
For values much above the maximum value of the coupling we consider,  
we find that with these binary parameters the neutron star becomes ill-behaved, suggesting we are 
approaching the value where no spherically symmetric neutron star exists for this EOS.
We estimate the initial orbital eccentricity to be $\sim 6 \times 10^{-3}$.
For the shorter inspiral, we consider an additional coupling of $\lambda/m_{\rm BH}^2=0.05$.

The second system we consider has binary parameters consistent with the recent
GW230529 event \cite{LIGOScientific:2024elc}. The source of GW230529 has component masses 
$3.6^{+0.8}_{-1.2}\ M_{\odot}$ and $1.4^{+0.6}_{-0.2}\ M_{\odot}$
and mass ratio
$q=0.39^{0.41}_{-0.12}$ at the $90\%$ confidence level. The primary spin most likely has a 
negative component when projected onto the orbital angular momentum, 
but is also consistent with zero spin: $\chi_1=0.44^{+0.40}_{-0.37}$.
The spin and tidal deformability of the neutron star were unconstrained, and 
no electromagnetic
counterpart has been identified to date.
Using the high-spin combined posterior samples, Ref.~\cite{LIGOScientific:2024elc}
found that the probability that the neutron star was tidally disrupted
is 0.1, corresponding to an upper limit on the remnant baryon mass produced
in the merger of $0.052 M_{\odot}$ at the $99\%$ confidence interval.
Yet this source is the most
probable of the BHNS events reported by the LVK to have undergone
tidal disruption because of the increased symmetry in the component masses. 
We therefore consider a non-rotating neutron star 
with gravitational mass $m_{\rm NS} = 1.4\ M_{\odot}$ 
and a non-spinning black hole with a mass of
$m_{\rm BH}=3.5\ M_{\odot}$, so that for the EOS we choose the neutron star is tidally
disrupted at merger.
The initial separation is
$D = 9.82 M$, or approximately 5 orbits before merging in GR. We consider
coupling values of $\lambda/m_{\rm BH}^2=$ 0,  0.1, and 0.15, where the maximum value
of the coupling here approaches limit where hyperbolicity of black hole solution
breaks down during scalarization process ($\lambda/m_{\rm BH}^2\approx 0.23$).

For ease of comparisons with other works, we convert our coupling $\lambda$ into
$\alpha_{\rm GB} \equiv \lambda/\sqrt{8\pi}$ used in, e.g.,
Refs.~\cite{Perkins:2021mhb,Lyu:2022gdr}\footnote{
 However, several other studies
 (e.g.~Refs.~\cite{Witek:2018dmd,Blazquez-Salcedo:2016enn,Pierini:2021jxd,Pierini:2022eim})
 take conventions leading to a value of $\alpha_{\rm GB}$
 that is $16\sqrt{\pi}\times$ times larger.
}.
Restoring physical units we have,
\begin{align}
   \sqrt{\alpha_{\mathrm GB}}
   \approx
   3.77 \
   \mathrm{km}
	\left(\frac{\sqrt{\lambda}}{m_{\rm BH}}\right)\left(\frac{m_{\rm BH}}{5.7\ M_{\odot}}\right)
    .
\end{align}
For reference, Ref.~\cite{Lyu:2022gdr} sets a constraint
of $\sqrt{\alpha_{\rm GB}} \lesssim 1.18$ km at a 90\% confidence level 
by comparing gravitational wave
observations of BHNS binaries to PN results for EsGB.
In comparison, the largest coupling we consider for the GW200115-like event 
($\lambda=0.1 m_{\rm BH}^2$) corresponds to $\sqrt{\alpha_{\rm GB}} \sim 1.19$ km,
which is at the limit of the observational bound. For the tidally disrupted
event, where the mass of the black hole is smaller,
the largest coupling ($\lambda=0.15 m_{\rm BH}^2$) corresponds to
$\sqrt{\alpha_{\rm GB}} \sim 0.89$ km, i.e. within the observational bounds.
We summarize the parameters of the BHNS systems we consider in Table~\ref{tab:summary}.
%------------------------------------------------------------------------------
\subsection{Diagnostic quantities\label{sec:diagnostics}}

We use many of the same diagnostics as in 
Refs.~\cite{East:2020hgw,Corman:2022xqg}, which we briefly review here.
We measure the scalar and gravitational
radiation by extracting the scalar field $\phi$ and the Newman-Penrose
scalar $\Psi_4$ on coordinate spheres at large radii ($r=100M$ where $M$ is the total mass).
We decompose
$\Psi_4$ and $\phi$ into their spin $-2$ and spin $0$ spherical harmonic components.
We use the average value of $\phi$ at large radius $r=100M$ to calculate the 
scalar charge $Q_{\rm SF}$.
We sometimes find it useful to consider the gravitational wave 
strain $h \equiv h_{+} + i h_{\times}$ instead, 
related to $\Psi_4$ through $\Psi_4 = \ddot{h}$. We numerically integrate
$\Psi_4$ using fixed frequency integration \cite{Reisswig:2010di}.

We track the apparent horizon associated with the black hole,
and measure its area and associated angular momentum $J_{\rm BH}$. 
From this, we compute the black hole mass $m_{\rm BH}$
via the Christodoulou formula \cite{Christodoulou:1970wf}.
We also track the total fluid rest mass outside the black hole horizon
\begin{equation}
	M_0 = \int \rho u^t \sqrt{-g} d^3 x
\end{equation}
where $\rho$ is the rest-mass density and $u^a$ is the four-velocity of the fluid.

\subsection{Post-Newtonian theory\label{sec:pn_theory}}
In this section, we summarize existing PN predictions for gravitational
and scalar waveforms in EsGB gravity. We perform a comparison with our numerical
waveforms in Sec.~\ref{sec:pn_comparison}.
As pointed out in Ref.~\cite{Sennett:2016klh}, the relative size of
the leading scalar dipolar radiation to the leading tensor quadrupolar radiation is
\begin{equation}
	\frac{\mathcal{F_{\rm nd}}}{\mathcal{F}_{\rm d}}= 
	\frac{24 x}{5 \zeta \mathcal{S}_{-}^2}
\end{equation}
with $\mathcal{F}$ denoting the energy flux rate and the subscripts `d' and `nd' denoting
the dipolar and non-dipolar part of the energy flux rate, respectively.
$x = (G_{\rm AB} M \Omega)^{\frac{2}{3}}$ is the PN expansion
parameter (see, e.g., Refs.~\cite{Sennett:2016klh,Julie:2022qux}) where
${G}_{\rm AB}=G\left(1+\alpha_A\alpha_B\right)$ is 
the effective gravitational constant
and $\alpha_{A/B}$ is the scalar charge
of body A(B), and the gravitational constant is reintroduced for
clarity. Finally, $\Omega$ is the orbital 
frequency, which we approximate as half the gravitational wave frequency
\cite{PhysRev.131.435,Berti:2007fi,Maggiore:2007ulw}. In addition, $\zeta$ and $\mathcal{S}_{-}$
are PN parameters that depend on theory considered and are 
summarized in Table I of Ref.~\cite{Julie:2022qux}.For the range of frequencies and 
coupling values we probe in our simulations, we find that the quadrupolar radiation
dominates over the dipolar radiation by a factor of $\sim 40-100$, meaning we are in the
so-called \emph{quadrupole driven regime} \cite{Sennett:2016klh}.

We first consider the gravitational modes $h_{\ell m}$ whose PN expression have been
computed to 2PN\footnote{We adopt the convention that all PN order are relative
to the quadrupolar radiation in GR. 
In this convention, the dipolar radiation enters at -0.5PN 
in the waveform and -1PN in the energy flux \cite{Bernard:2022noq}.} 
order in scalar-tensor theories \cite{Sennett:2016klh}
and, more recently, to 1PN directly in EsGB~\cite{Shiralilou:2020gah,Shiralilou:2021mfl}.
We consider the PN expressions from Ref.~\cite{Sennett:2016klh}
\begin{equation}\label{eq:PN_GW}
	\frac{r}{M} h_{\ell m} = 2 \tilde{G} (1-\zeta)\eta x \sqrt{\frac{16 \pi}{5}}
	\hat{H}_{\ell m} e^{-i m \psi},
\end{equation}
where $\eta= m_{\rm BH} m_{\rm NS}/M^2$ is the symmetric mass ratio, $\psi$ the orbital
phase given by Eqs.(60-61) of Ref.~\cite{Sennett:2016klh}, 
and $\tilde{G}$\footnote{Comparing with Eq.~(65) of \cite{Sennett:2016klh}
we note that we have replaced $G$ with $\tilde{G}$ to avoid confusion with
our gravitational constant $G$. $\tilde{G}$ is the notation used in Ref.~\cite{Bernard:2022noq}
and Table I of Ref.~\cite{Julie:2022qux}.} is a PN parameter again defined in Table I of 
\cite{Julie:2022qux}. The expressions for the amplitude modes $\hat{H}_{\ell m}$ are long,
and given in Eq.~(67) of Ref.~\cite{Sennett:2016klh}. We have mapped these expressions 
to EsGB using the mapping outlined in Sec.~IV.A of Ref.~\cite{Pompili:2024}. 
Note that, tidal effects, which enter into the phase evolution at 5PN \cite{Flanagan:2007ix},
were ignored in Ref.~\cite{Sennett:2016klh}. This is a reasonable assumption here, 
since after using the values listed in first row of Table~\ref{tab:summary} we find
that
the mass-weighted tidal deformability $\tilde{\Lambda}^{\rm GR} \sim 13$ of the system
is small, and hence is expected to have little impact on the binary dynamics.
We also note that scalar-induced dipolar tidal effects derived to leading order 
for nonspinning binary black holes in EsGB \cite{vanGemeren:2023rhh}, 
and to next-to-next-to-leading order in scalar-tensor
theories \cite{Bernard:2023eul}, vanish for shift-symmetric EsGB gravity.

We next consider the spherical harmonic components
of the scalar radiation $\phi_{\ell m}$. 
These were derived to relative 0.5PN order in EsGB in 
Refs.~\cite{Shiralilou:2020gah,Shiralilou:2021mfl},
and relative 1.5PN order (2PN order beyond the leading dipolar contribution in waveform)
in scalar tensor theories by 
Ref.~\cite{Bernard:2022noq}. 
Here, we use the results of Ref.~\cite{Bernard:2022noq} and map them
to EsGB using the mapping of Sec.IV.B of Ref.~\cite{Pompili:2024}, keeping only 
leading order terms in $\lambda/m_{\rm BH}^2$. The expressions can be found in Appendix E
of Ref.~\cite{Pompili:2024}\footnote{Note that Ref.~\cite{Pompili:2024} uses different
conventions from this paper so that, $\phi = \sqrt{2} \varphi$ where $\varphi$
is scalar field in Ref.~\cite{Pompili:2024}.}.

%=============================================================================
\section{Results\label{sec:results}}
We follow the evolution of two types 
of BHNS binaries distinguished by whether the neutron
star is tidally disrupted before merger (see Table~\ref{tab:summary}). 
For both scenarios, we vary the EsGB coupling
all the way up to near the maximum value for which we were able to carry out the
evolution. We first consider a system with GW200115-like parameters and evolve it 
both in GR and EsGB gravity with a coupling comparable to the 
upper bound obtained in Ref.~\cite{Lyu:2022gdr}.
In Sec.~\ref{sec:inspiral}, we first focus on the dynamics during inspiral 
and show that the EsGB system inspiral faster
than its GR counterpart. We then compare both the scalar and gravitational radiation to
predictions from PN theory in Sec.~\ref{sec:pn_comparison}. Our main result is that the
PN prediction is a good approximation to the amount of dephasing in binary 
up to late in the inspiral.
In Sec.~\ref{sec:merger}, we focus on the merger dynamics and trends with varying EsGB coupling.
We find a negligible change in amplitude of gravitational wave signal at merger.
Independently of whether the neutron star is tidally disrupted, the main effect on the ringdown signal
is not a shift in ringdown frequencies but
a change in the amplitude of the signal. Specifically, the amplitude increases (or decreases) with increasing coupling 
when the neutron is (or is not) tidally disrupted. 
In the case where the neutron star is tidally disrupted, we also study the
amount of material remaining outside the black hole after the merger and find a 
slight decrease in the amount of material
leftover with increasing coupling due to the neutron star being more compact.

%=============================================================================
\subsection{Comparison between GR and EsGB during inspiral}\label{sec:inspiral}

We first consider the BHNS system with GW200115-like parameters and an initial separation
of $D = 10.35 M$. We evolve the system both in GR and 
with a coupling value of $\sqrt{\alpha_{\rm GB}} = 1.19\ \rm{km}$.
For the parameters and EOS we consider, the neutron star is swallowed by the black hole
without tidal disruption (see the first row of Table~\ref{tab:summary}). 

\begin{figure}
	\includegraphics[width=0.99\columnwidth,draft=false]{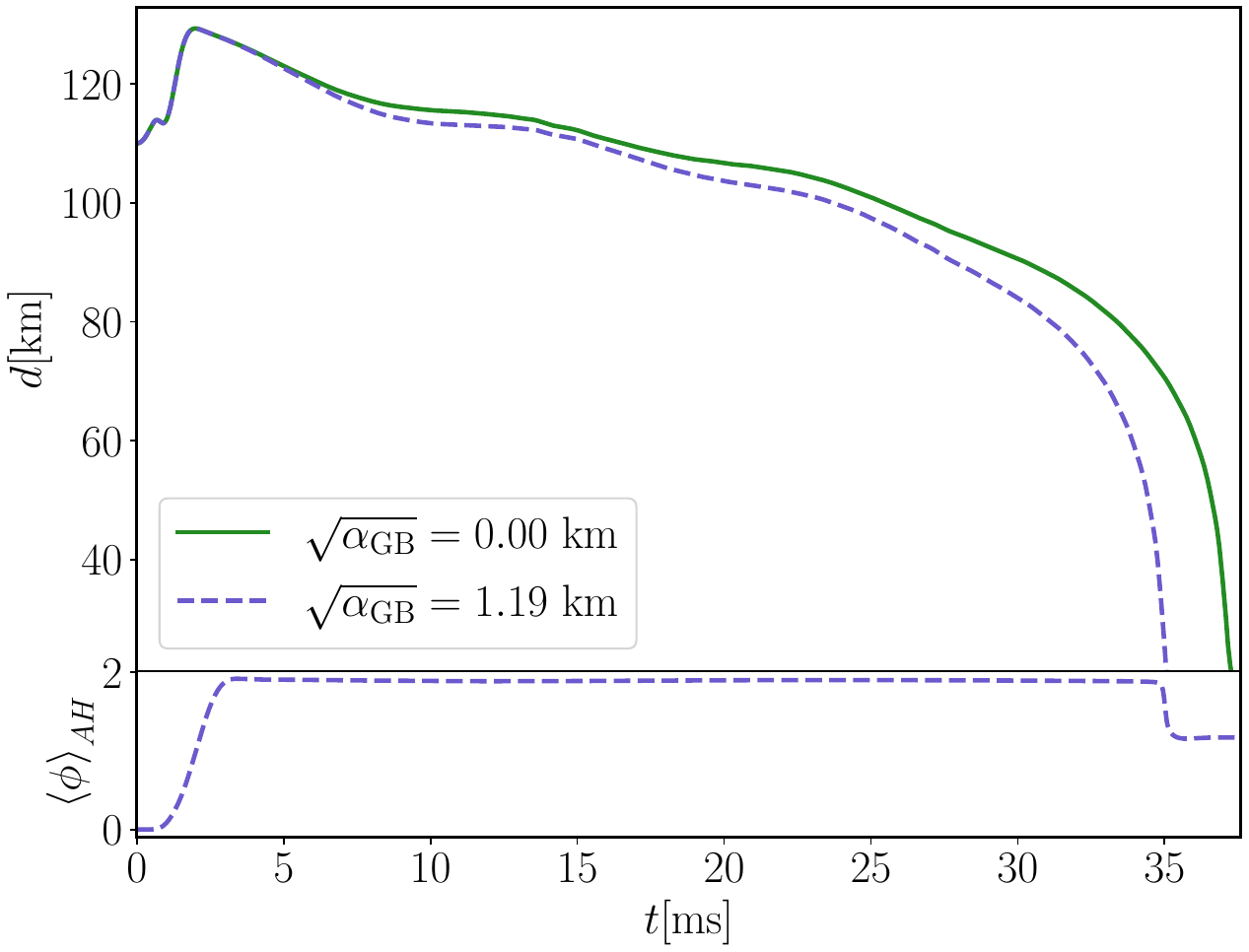}
        \caption{(Top) Coordinate separation of BHNS merger with binary
        parameters consistent with the GW200115 event in GR and EsGB gravity with a
        coupling value of $\sqrt{\alpha_{\rm GB}} = 1.19\ \rm{km}$. With the chosen
        initial separation, the binary undergoes $\sim 7$ orbits before merger in
        GR. The increase in coordinate separation during the 
	first $2$ ms is due to the transition to damped harmonic gauge. 
	The bottom panel shows the value of the scalar field averaged on the
        black hole apparent horizon for the EsGB case.
\label{fig:GW200115_long_traj}
}
\end{figure}

For simplicity, we use the same initial data for the EsGB system and its GR counterpart,
i.e. we set $\phi=\partial_t \phi = 0$ on the initial time slice but slowly
ramp on the coupling of the theory over $100M$. 
In the bottom panel of Fig.~\ref{fig:GW200115_long_traj}, 
we show the average value of the scalar field
on the black hole apparent horizon.
Note that the black hole acquires its scalar charge on a timescale much shorter
than the inspiral timescale and 
we therefore do not expect our results after this transitory
period to be noticeably affected. We have also checked that the scalarization process
does not appreciably impact the orbital eccentricity or increase the level of constraint
violation compared to that coming from truncation error (see Appendix~\ref{app:convergence}).

The top panel of Fig.~\ref{fig:GW200115_long_traj} shows the evolution of the coordinate
separation between the two compact objects for the GR and EsGB systems. We first note
that the merger part of both systems can be aligned through a time shift (see below). This
implies they have a similar orbital separation or frequency for the onset of the plunge.
This arises because the gravitational attraction in EsGB gravity is characterized by
the effective gravitational constant ${G}_{\rm AB}=G\left(1+\alpha_A\alpha_B\right)$
introduced earlier.
Since the neutron star charge
$\alpha_{\rm NS}= \mathcal{O}(\lambda^3)$ is negligible, the gravitational pull
of a BHNS system in GR and EsGB is similar. By contrast, in binary black hole mergers,
both black holes carry a positive scalar charge (see below for the explicit value), 
which increases the gravitational pull
and therefore the orbital separation at which the objects merge.
However, the BHNS system in EsGB still admits an additional energy dissipation channel
via scalar radiation, and hence inspirals faster compared to GR, as can be seen from 
Fig.~\ref{fig:GW200115_long_traj}.

\begin{figure*}
	\includegraphics[width=0.99\columnwidth,draft=false]{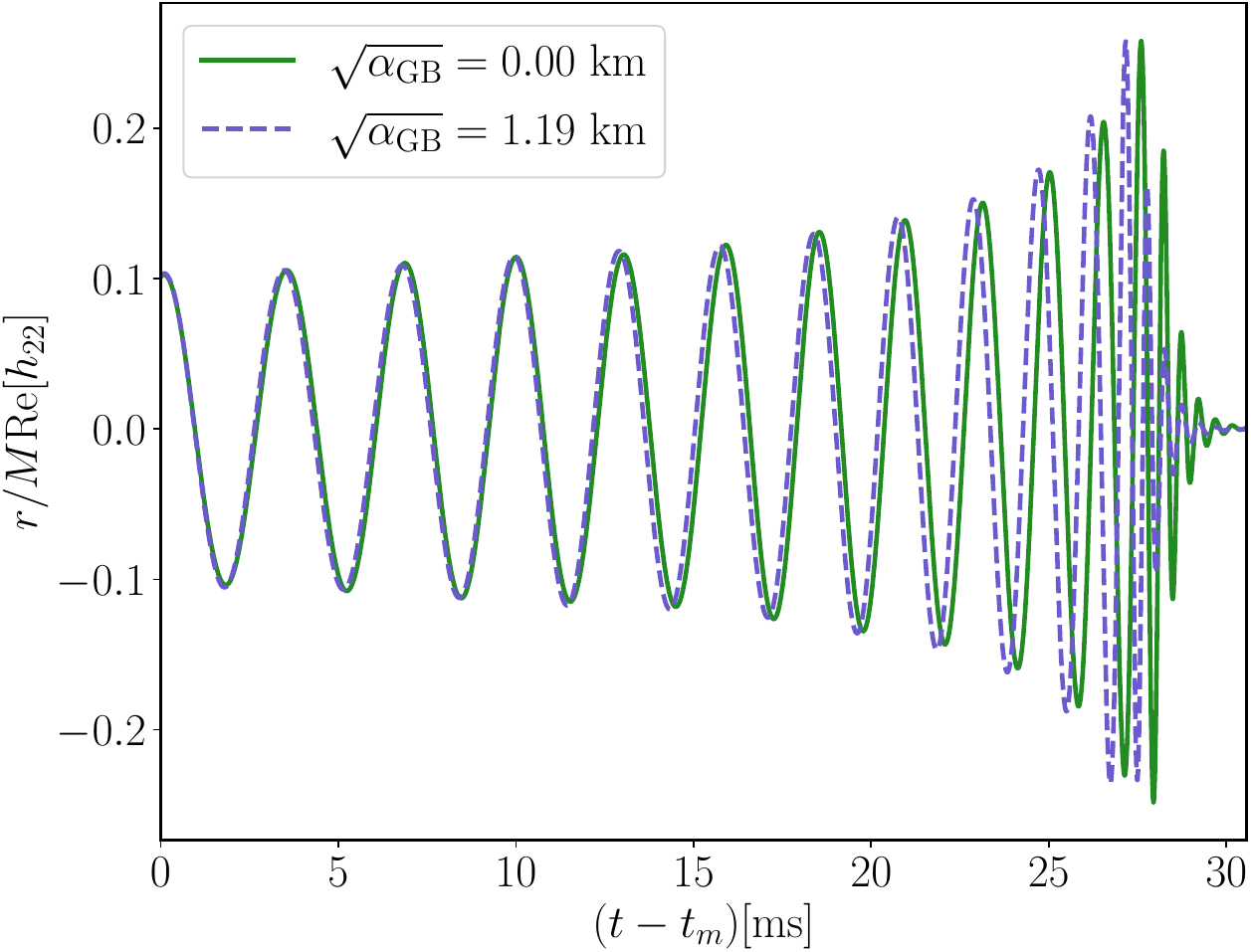}
	\includegraphics[width=0.99\columnwidth,draft=false]{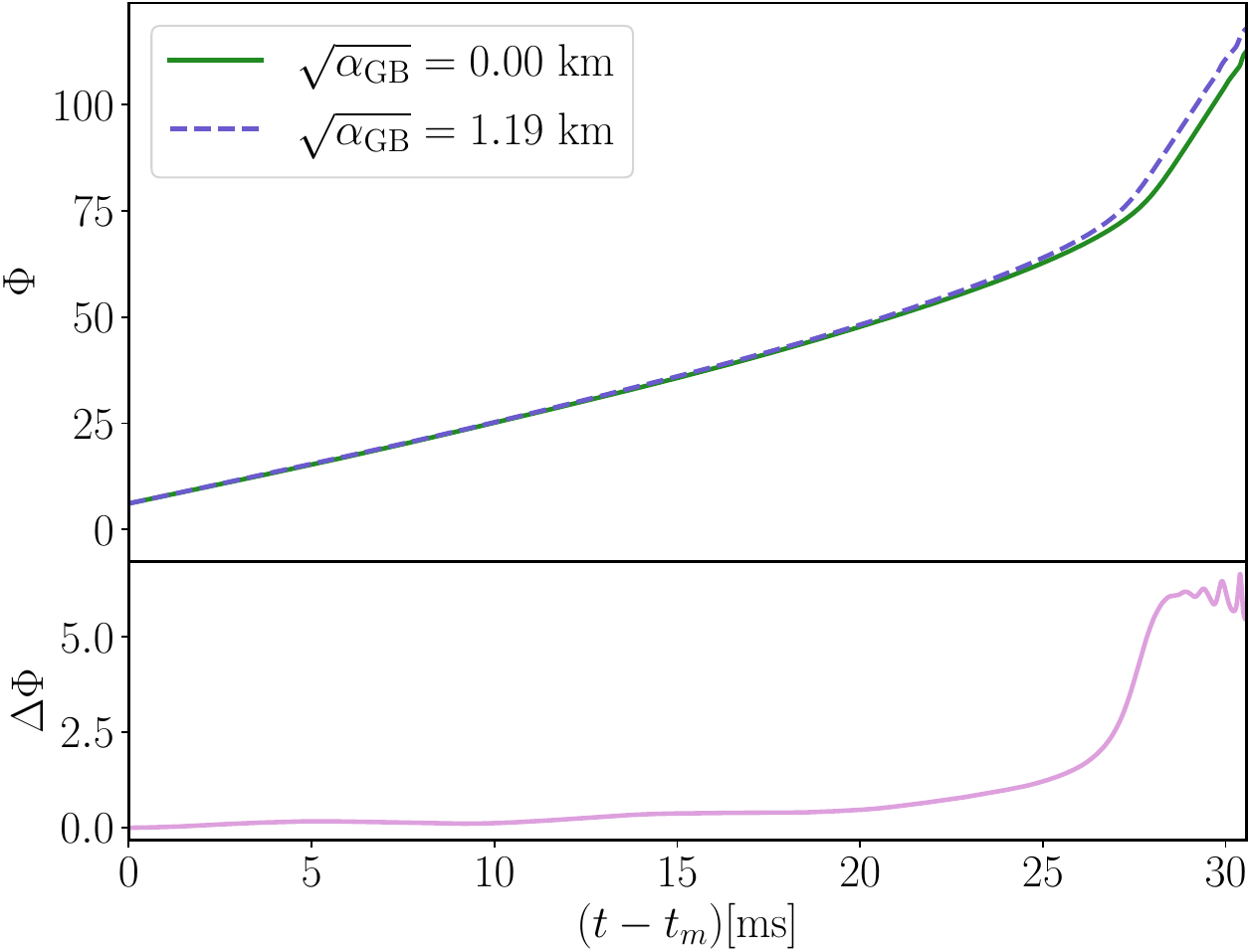}
	\caption{(Left) Gravitational wave radiation for 
		a BHNS merger with binary parameters consistent
		with the GW200115 event in GR and EsGB gravity with a coupling value of 
		$\sqrt{\alpha_{\rm GB}} = 1.19\ \rm{km}$.
		We show the real part of the $\ell = m = 2$
		spherical harmonic of the gravitational wave strain $h$
		extracted at $100M$. (Right) The 
		gravitational wave phase of the aligned waveforms $\Phi$. 
		The bottom shows the phase difference between the two waveforms
		$\Delta \Phi = \Phi_{\rm GB}-\Phi_{\rm GR}$.
                The waveforms have been aligned in time and phase at a
		gravitational wave frequency $f_m=0.01/M$, or equivalently
		$t_m = \{9.7,8.1 \}\rm{ms}$ for 
		$\sqrt{\alpha_{\rm GB}} = \{0.0,1.19\} \rm{km}$ according to
		Eq.~\ref{eq:align_h}.
\label{fig:GW200115_long_gw}
}
\end{figure*}

The shorter inspiral in EsGB gravity also leads to a shorter gravitational wave signal.
Figure~\ref{fig:GW200115_long_gw} shows the $\ell=m=2$ harmonic of the strain
in GR and EsGB. We align the two waveforms by requiring that the EsGB and GR waveform
agree in time and phase at some fiducial frequency $\omega_m$. More precisely, we leave the GR waveform
untouched but construct a new shifted EsGB waveform:
\begin{equation}\label{eq:align_h}
	h_{22}^{\rm{GB}'}(t) = h_{22}^{\rm{GB}}(t+t_c-t_m) e^{i \left(\Phi_{\rm{GB}}(t_c)-\Phi_{\rm{GR}}(t_m)\right)},
\end{equation}
where $t_c$ is the time so that the derivative of the complex phase of EsGB waveform satisfies
$\dot{\Phi}_{\rm GB} (t_c) = \omega_m$ and similarly $t_m$ is the time where
$\dot{\Phi}_{\rm GR}(t_m) = \omega_m$. Note that the gravitational wave
frequency computed from the
time derivative of the complex phase of the numerical waveform is typically noisy at early
times and becomes smoother near the merger. To allow a matching at any time, we fit a
polynomial in time through the frequency.
In Fig.~\ref{fig:GW200115_long_gw}, $M f_m=M \omega_m/(2\pi)$ 
was chosen to be $0.01$.
We also show the phase evolution $\Phi$ of the aligned waveforms in the right panel
of Fig.~\ref{fig:GW200115_long_gw}, as well as the corresponding waveform phase
differences,
\begin{equation}\label{eq:delta_phi_nr}
	\Delta \Phi = \Phi_{\rm GB} - \Phi_{\rm GR}
\end{equation}
in the bottom right panel of Fig.~\ref{fig:GW200115_long_gw}.
After the amplitude of the EsGB
waveform peaks, it takes the GR waveform another $\sim 4.3$ radians to peak.
Similarly to the
binary black hole mergers in Ref.~\cite{Corman:2022xqg}, we find that the
dominant truncation error in our simulations does not depend strongly on the
value of the coupling and therefore partially cancels out when calculating the
difference in gravitational wave phase between EsGB and GR simulations using
the same resolution. See Appendix~\ref{app:convergence} for details.
We estimate the truncation error in $\Delta \Phi$ to be $\sim 1.2$. This is smaller
than the estimated phase error in the GR waveform itself.

\subsection{Comparison to post-Newtonian theory}\label{sec:pn_comparison}
We now quantitatively compare the gravitational 
and scalar waveforms
we obtain from our numerical evolution with the existing PN predictions in EsGB gravity
outlined in Sec.~\ref{sec:pn_theory}.
Along the way, we comment on the accuracy of current methods using PN predictions to constrain
EsGB gravity.
\begin{figure*}
	\includegraphics[width=0.99\columnwidth,draft=false]{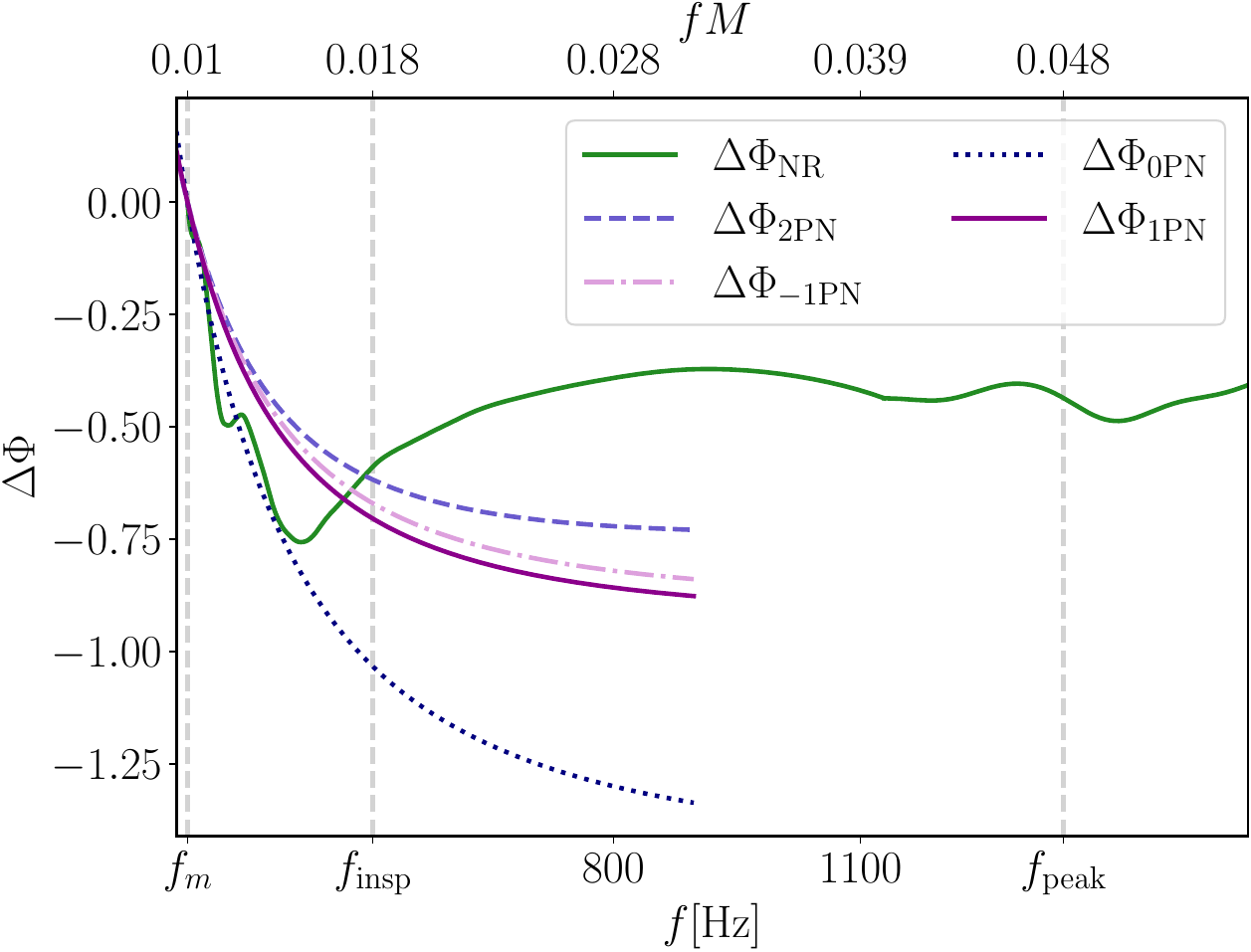}
	\includegraphics[width=0.99\columnwidth,draft=false]{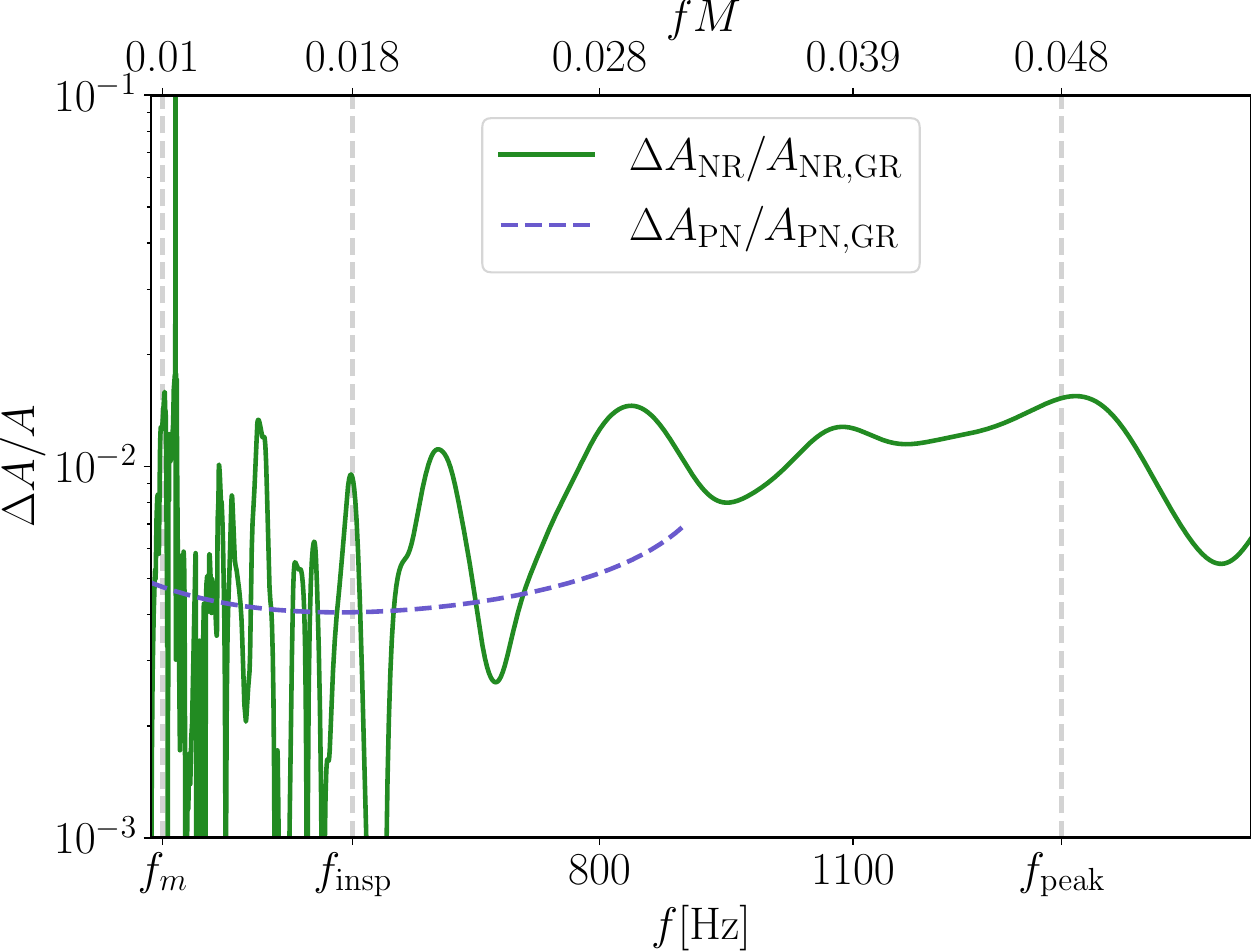}
	\caption{PN and numerical predictions for the dephasing (left) 
	and correction to the amplitude (right) of gravitational waveform of
	the binary shown in Fig.~\ref{fig:GW200115_long_gw} (see Eq.\ref{eq:delta_phi_nr},
	\ref{eq:dephasing_pn},
	\ref{eq:dephasing_pn}
	and \ref{eq:amplitude_diff}). 
	We also show the dephasing when only considering leading
	order (dipolar) contribution to phase and all contributions up
	to 0PN and 1PN order. The first vertical line corresponds to frequency at which
	waveforms are aligned $f_{m}$ according to Eq.\ref{eq:align_h}, the second line
	to the
	end of inspiral stage in GR $f_{\rm insp}$, and the third
	to frequency at which amplitude of GR waveform peaks
	$f_{\rm peak}$.
\label{fig:GW200225_PN_GW}
}
\end{figure*}

Considering the same system as in the previous section,
we use Eq.~\ref{eq:PN_GW} to compute the PN prediction for dephasing of gravitational
wave phase,
\begin{equation}\label{eq:dephasing_pn}
	\Delta \Phi_{\rm PN} = 2 \Delta \psi = 2 \left(\psi_{\rm{GB}}-\psi_{\rm GR}\right)
\end{equation}
and compare this to the
dephasing in our numerical simulation, $\Delta \Phi_{\rm NR}$ 
(see Eq.~\ref{eq:delta_phi_nr} and Fig.~\ref{fig:GW200115_long_gw}). Similarly we
also compare the relative change in amplitude of waveform, 
\begin{equation}\label{eq:amplitude_diff}
	\frac{\Delta A}{A} = \frac{|h^{\rm GB}_{22}|-|h^{\rm GR}_{22}|}{|h^{\rm GR}_{22}|}
\end{equation}
computed using Eq.~\ref{eq:PN_GW} and in our numerical simulation.
In Fig.~\ref{fig:GW200225_PN_GW}, we show the numerical and PN prediction
for the relative change in amplitude 
and dephasing as a function of gravitational
wave frequency. We note that, as was mentioned in the previous section, although 
the binary in EsGB inspirals faster than in GR, the frequency at which the objects 
merge is similar, with an agreement of $\sim 2\%$ for this particular system.
More precisely, the left panel
shows the total
dephasing computed to 2PN, $\Delta \Phi_{\rm 2PN}$, but also 
when considering the leading dipolar (-1PN)
contribution only, $\Delta \Phi_{\rm -1PN}$, as well as the 
dephasing computed up to 0PN and 1PN (we omit the
0.5PN and 1.5PN results for clarity).
Since the EsGB corrections to the waveform within the
PN expansions are only expected to be valid during the inspiral stage,
tests of GR on the inspiral turn off corrections to GR at some cutoff frequency,
which was taken to be $f = 0.018/M$ (or $\sim 500$ Hz
for this particular system) in Ref.~\cite{Lyu:2022gdr}\footnote{Parametric tests 
on the inspiral done in the LVK analyses in Refs.~\cite{TheLIGOScientific:2016src,Abbott:2018lct,
LIGOScientific:2019fpa,LIGOScientific:2020tif,LIGOScientific:2021sio}
use a cutoff frequency
of $f_c^{\rm {PAR}}=0.35 f^{22}_{\rm peak} $ where $f^{22}_{\rm peak}$ is the
GW frequency at the peak amplitude of $(\ell=2,m=2)$ waveform. In our particular
setup, this would correspond to a frequency of $f M \sim 0.017$.}.
We find that the difference between the PN and numerical results 
up to $f = 0.018/M$ are smaller than
or comparable to our estimates of the truncation error in $\Delta \Phi$.
Our result suggests that using the PN expansion up to $f = 0.018/M$ 
to constrain EsGB gravity, as was done
in Ref.~\cite{Lyu:2022gdr}, is a good approximation.

A similar argument would apply to constraints that would be obtained by mapping
constraints on the -1PN coefficient (from a parameterized test)
to constraints on $\sqrt{\alpha_{\rm GB}}$
through the post-Einsteinian formalism \cite{Yunes:2009ke},
as was done in, for instance, Refs.~\cite{Yunes:2016jcc,Sanger:2024}.
It would be interesting to perform parameter estimation on our numerical
simulations using not only the PN results for EsGB, 
as was done for the observational data in Refs.~\cite{Lyu:2022gdr,Gao:2024rel};
but also using theory-agnostic approaches, such as the TIGER
\cite{Agathos:2013upa,Meidam:2017dgf} or FTI
\cite{Mehta:2022pcn} frameworks,
which constrain the PN coefficients by varying them one at a time. 
However, this is beyond the scope of 
this paper, and we leave it to future work.
Note that the left panel of Fig.~\ref{fig:GW200225_PN_GW} also shows that the leading
order dipolar contribution to the dephasing dominates.
Although a more detailed study would be required,
this indicates that constraints on the leading 
PN coefficient recovered when only
variations at that particular order are allowed (as is typically done
in most current analyses), 
would in the case of EsGB, be a 
satisfactory assumption, which was already argued in Ref.~\cite{Perkins:2022fhr}.

We note that the PN prediction combining all orders predicts
that the EsGB system should inspiral faster than GR at any given frequency all
the way up to near merger, 
but that this effect diminishes at high frequencies. Moreover, we find that the
highest PN corrections, namely 1.5 and 2PN (see Fig.~\ref{fig:GW200225_PN_GW}), 
reduce the rate of inspiral with respect to
GR. This trend is consistent with our numerical simulations which show 
in Fig.~\ref{fig:GW200225_PN_GW} that the extra dephasing past the inspiral stage
is negligible. This is in agreement with Ref.~\cite{Julie:2022qux}, where it
was found that when considering the conservative part of 
the dynamics, including higher PN
orders tends to decrease the orbital frequency of the binary at the innermost stable circular
orbit.
This further suggests that setting the corrections to the phase to zero
past the inspiral stage, as in Refs.~\cite{Lyu:2022gdr,Sanger:2024,Gao:2024rel} 
is a good approximation,
at least for EsGB gravity where we find that the changes to the phase during
merger and ringdown are small.

Finally, we note that the PN prediction for the amplitude of waveform, shown in
right panel of Fig.~\ref{fig:GW200225_PN_GW}, is consistent with our numerical
results, and that the relative change in amplitude of waveform remains small throughout
the inspiral. This further motivates generic inspiral tests of GR 
on the phase rather than the amplitude of waveform.

\begin{figure*}
	\includegraphics[width=0.99\columnwidth,draft=false]{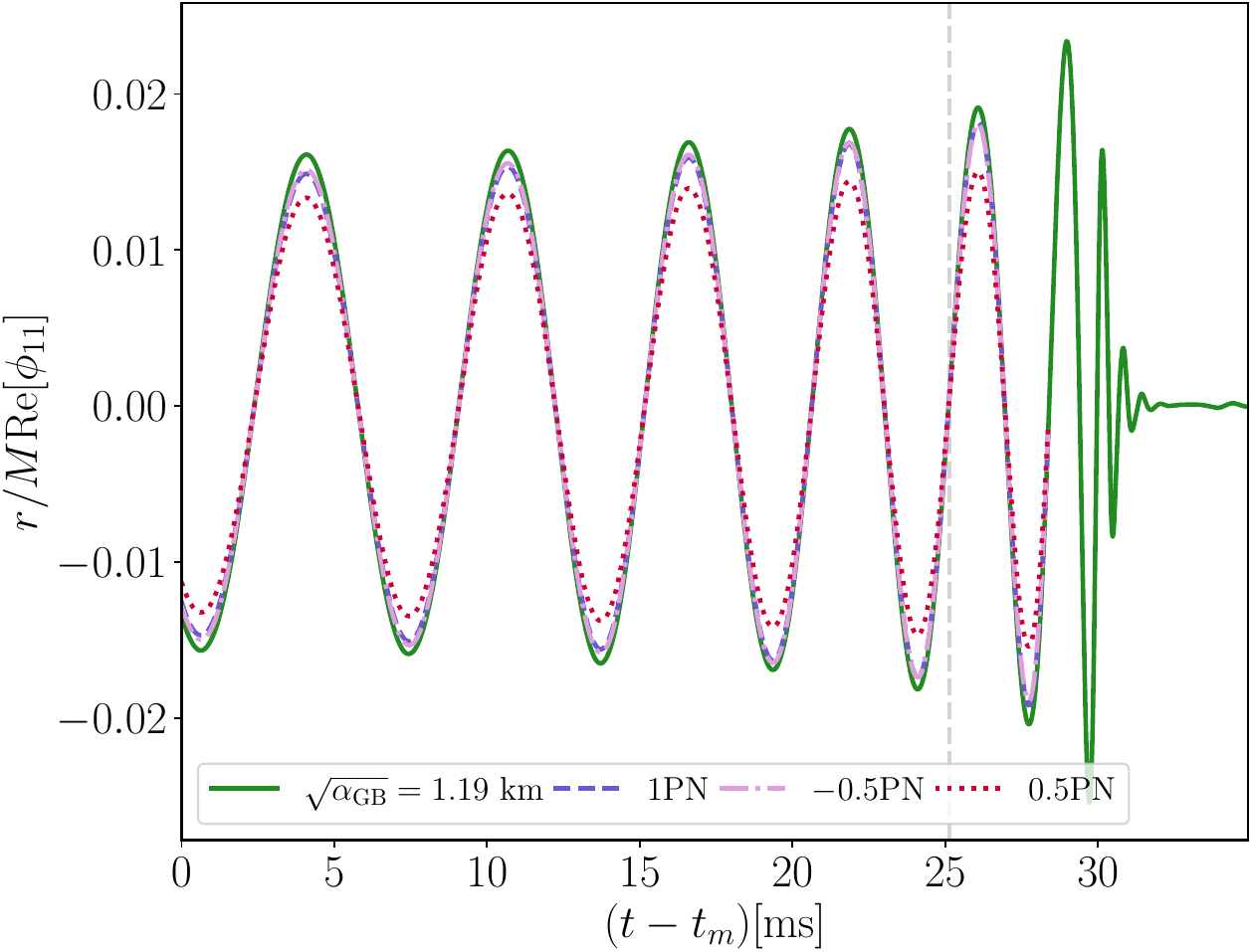}
	\includegraphics[width=0.99\columnwidth,draft=false]{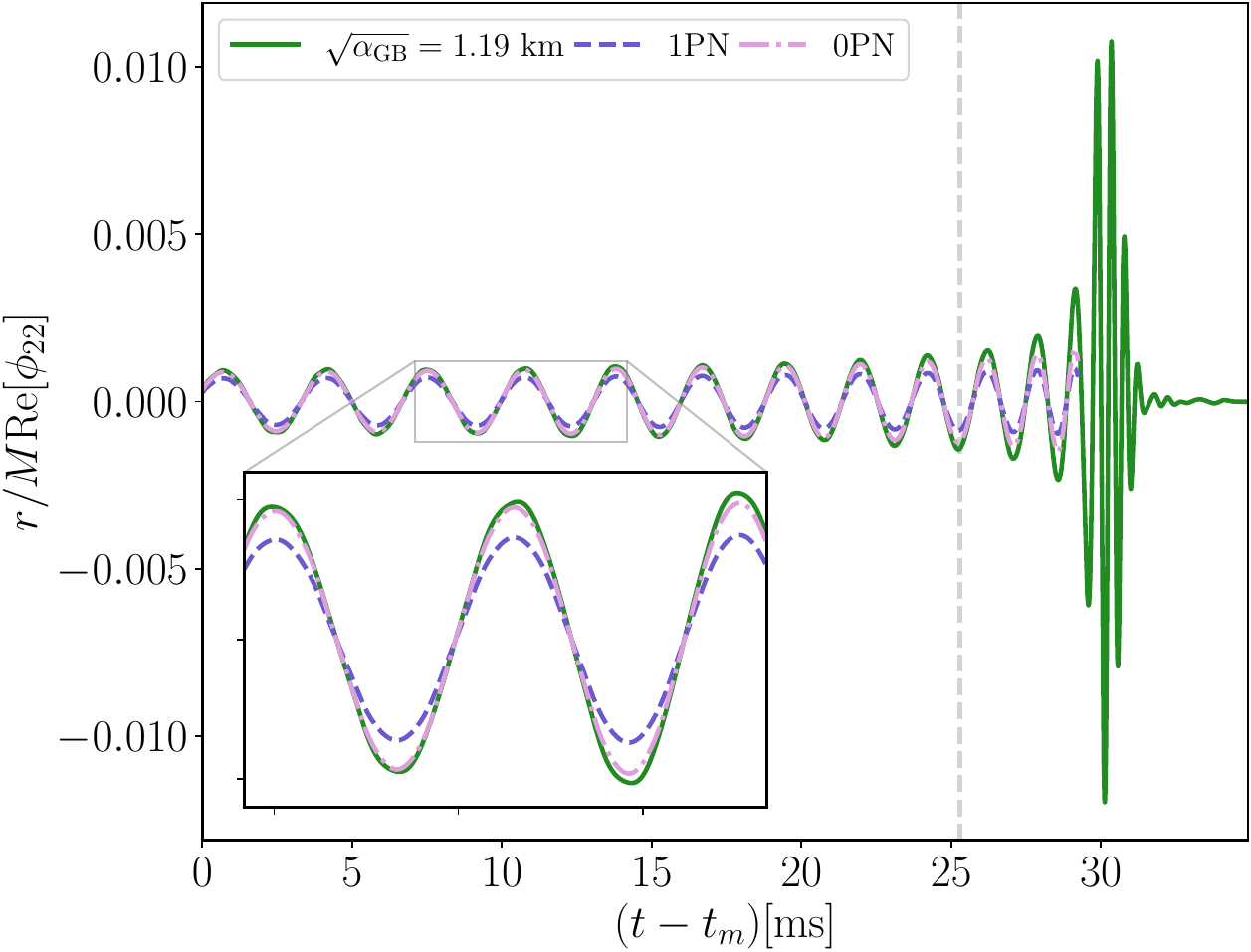}
        \caption{ 
		Scalar radiation for a 
		BHNS binary merger with binary parameters consistent
		with the GW200115 event for a coupling value of 
		$\sqrt{\alpha_{\rm GB}} = 1.19\ \rm{km}$.
		We show the real part of
		the $(\ell = 2,m = 2)$ (right) and $(\ell = 1,m = 1)$ (left)
		spherical harmonic of scalar waveform 
		$\phi$ extracted at $100M$.
		During the inspiral, we also display the PN predictions derived to 
		relative 
		1.5PN. The vertical line roughly corresponds to time at which modifications
		to GR are turned off in Ref.~\cite{Lyu:2022gdr}, i.e. 
		$t(f_{\rm insp}=0.018/M)$.
		Time is measured with respect to time where
                both waveforms have a gravitational wave frequency $f_m=0.01/M$.
\label{fig:GW200225_PN_scalar}
}
\end{figure*}

We next compare the spherical harmonic components
of the scalar radiation $\phi_{\ell m}$ extracted from our simulations with PN
predictions. In Fig.~\ref{fig:GW200225_PN_scalar}, we compare our numerical
scalar modes $\phi_{11}$ and $\phi_{22}$ to the PN predictions at 
successively higher PN orders.
As in the comparisons of scalar waveforms computed in 
Refs.~\cite{Witek:2018dmd,Shiralilou:2020gah,Shiralilou:2021mfl,Corman:2022xqg},
the frequency we use in the PN expressions 
are obtained from our numerical evolutions, 
so our comparison is measuring the accuracy of the PN
approximation in determining the amplitude of the scalar field, given
its frequency. We see that the lowest PN order 
contributes the most to the amplitude of waveform.
We find that the fractional difference between the numerical waveforms and the 1PN 
$(\ell=1,m=1)$ mode
is about $5\%$, while for the $(\ell=2,m=2)$ it is initially $\sim 22 \%$ and grows as
binary inspirals. We note that the inclusion of the next-to-leading order in the 
$(\ell=2,m=2)$ waveform worsens the agreement between the PN and numerical waveform,
so higher PN terms may be needed for better agreement with numerical simulations.

\subsection{Merger dynamics and trends with varying EsGB coupling \label{sec:merger}}
Lastly, we study the effects of EsGB on the merger and ringdown of two different scenarios:
a BHNS merger 
with the same intrinsic parameters as in the previous two sections, i.e. GW200115-like,
but with smaller initial separation and considering additional values for the coupling, 
and a GW230529-like BHNS merger where the neutron
is tidally disrupted before merger (see Table~\ref{tab:summary}). 
For these two systems, we vary the EsGB coupling
up to near the maximum value where we are able to carry out the evolution,
allowing us to study the trends with the coupling of the theory.

\begin{figure*}
	\includegraphics[width=0.99\columnwidth,draft=false]{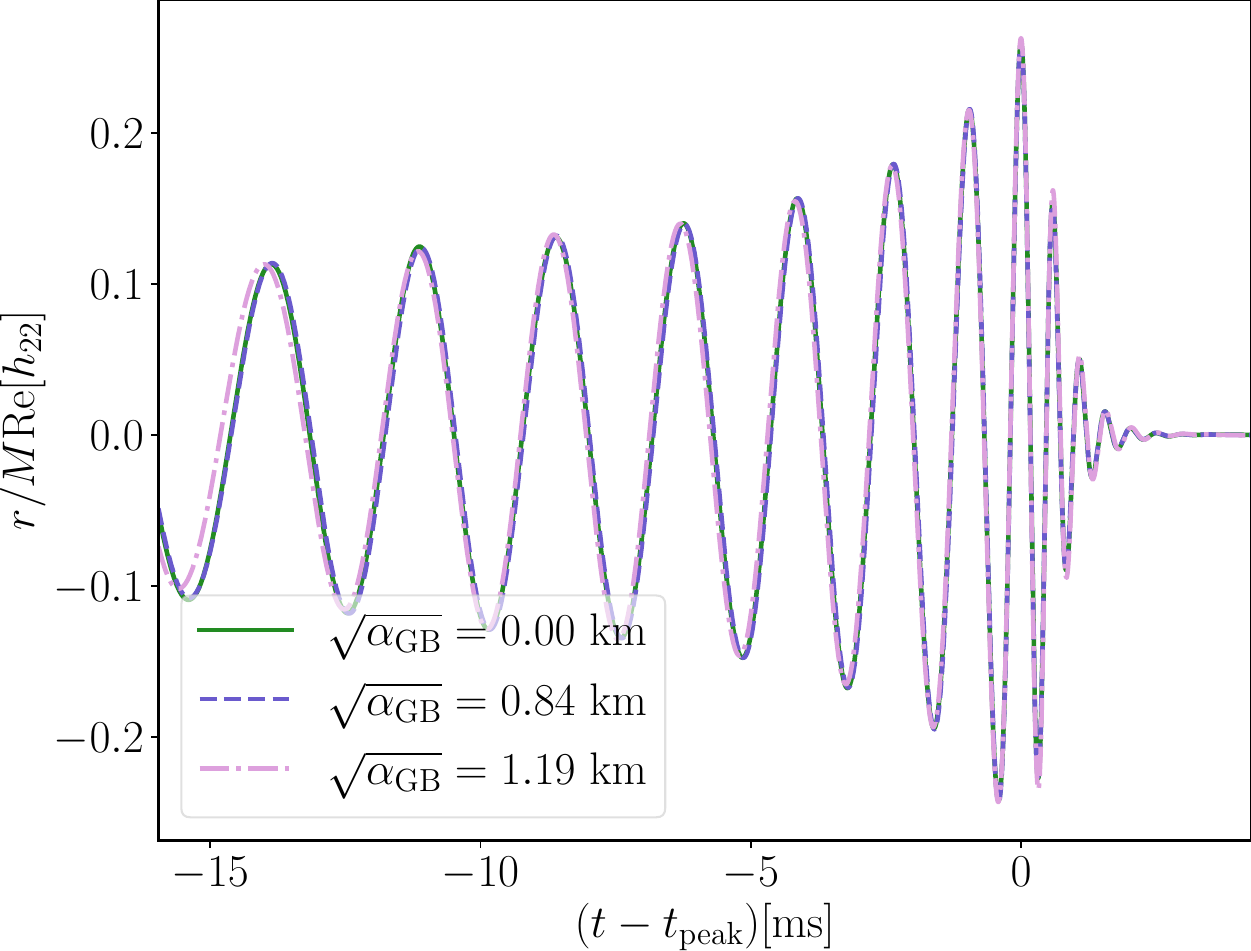}
	\includegraphics[width=0.99\columnwidth,draft=false]{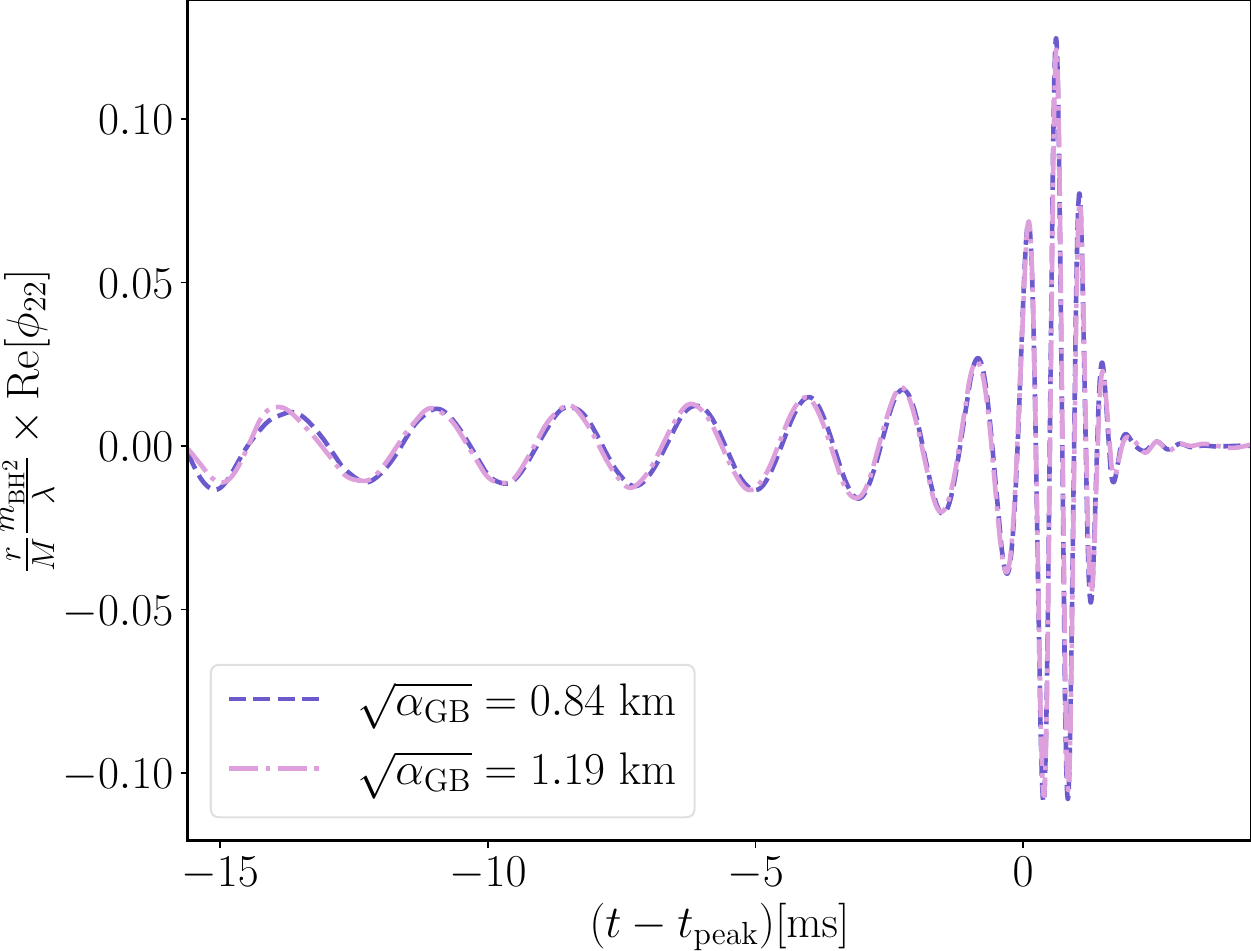}
        \caption{
		Gravitational wave radiation (left) and 
		scalar radiation (right) for a 
		BHNS merger with binary parameters consistent
		with the GW200115 event for different values of the EsGB coupling. 
		We show the real part of
		the $(\ell = 2,m = 2)$ spherical harmonic of the strain 
		$h$
		and $(2,2)$ component of $\phi$ both extracted at $100 M$. Time is 
		measured with respect to the time where amplitude of $h_{22}$
		is maximum $t_{\rm peak}$.
\label{fig:GW200225_short_wave}
}
\end{figure*}

Figure~\ref{fig:GW200225_short_wave} shows the gravitational (left) and scalar
radiation (right) for the first scenario. After $\sim 4-5$ orbits, the black hole swallows
the neutron star and rings down to form a final remnant black hole with a larger mass.
In addition to GR, we consider two other couplings, $\sqrt{\alpha_{\rm GB}} = 0.84\ \rm{km}$
and $1.19\ \rm{km}$, with the largest value at the limit of the observational bounds, 
$\sqrt{\alpha_{\rm GB}}\sim 1.19\ \rm{km}$.
Here we align the waveforms in time and phase at the
peak of amplitude of the gravitational waveforms. Despite the shorter inspiral,
we observe some dephasing in the $\ell=2,\ m=2$ harmonic of the strain (left panel),
consistent with EsGB binaries merging faster the larger their coupling
due to the additional energy loss through scalar radiation. The
$\ell=2,\ m=2$ component of the scalar radiation (right panel) shows similar
behavior to the gravitational waves both in the inspiral and ringdown.
After rescaling for the test-field dependence on coupling, we find that there is
a mild nonlinear enhancement just before merger, 
but this is negligible earlier in the inspiral 
and during ringdown.

\begin{figure*}
	\includegraphics[width=0.99\columnwidth,draft=false]{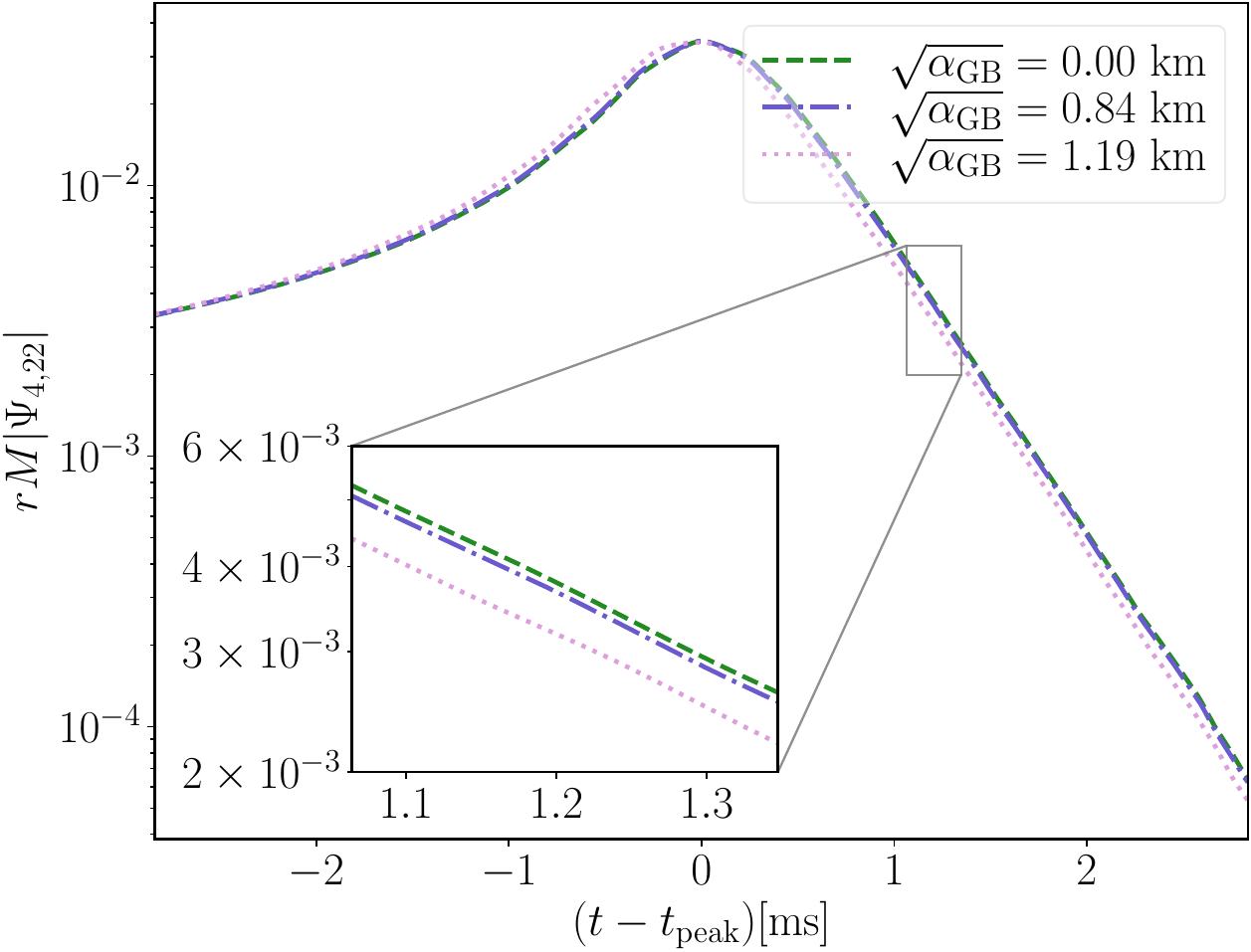}
	\includegraphics[width=0.99\columnwidth,draft=false]{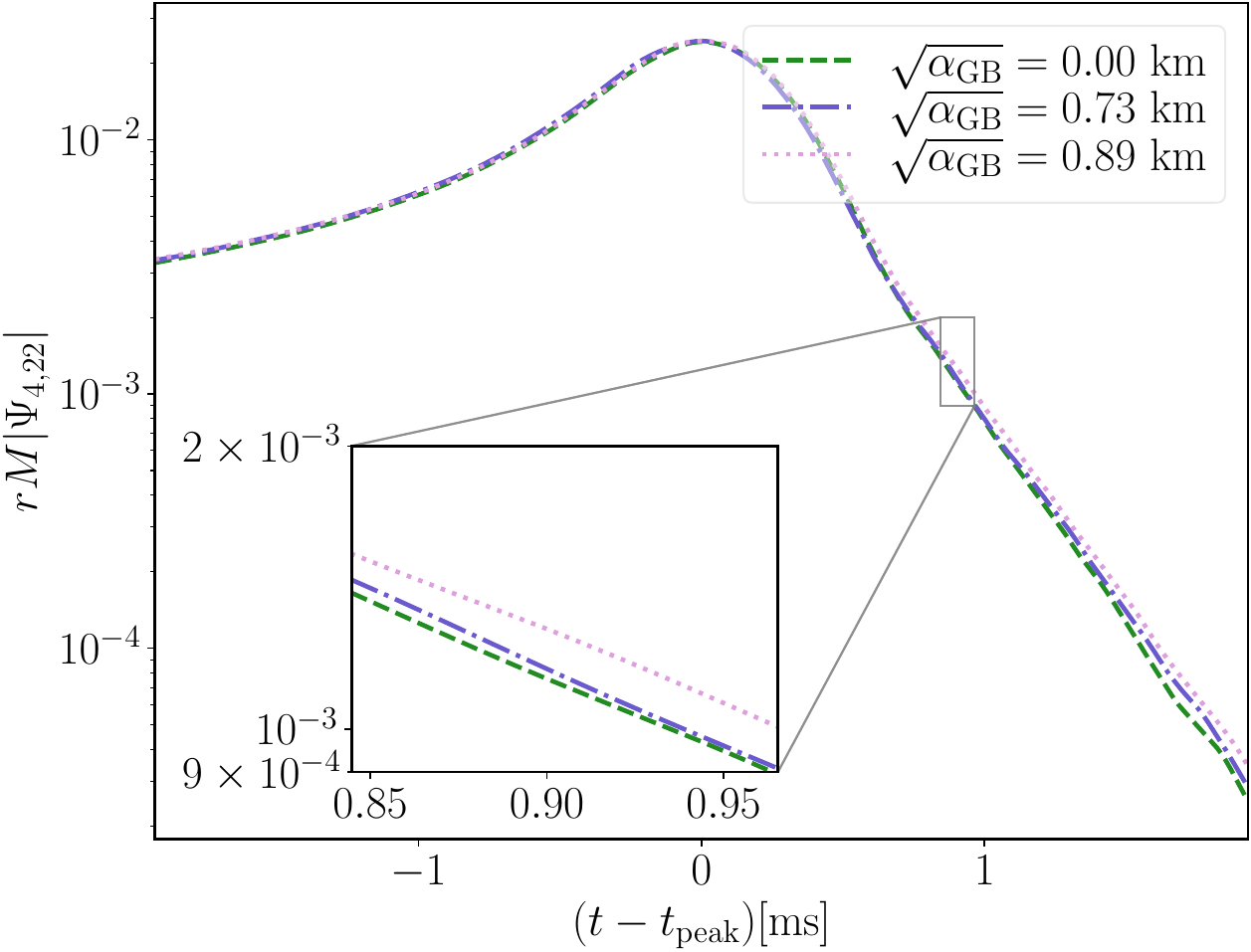}
        \caption{ The amplitude of the $\ell = 2,\ m=2$ spherical harmonic $\Psi_4$ 
	for a 
		BHNS binary merger with parameters consistent 
		with the GW200115 event (left) and for a binary where the neutron star is tidally
		disrupted (right) for different values of the EsGB coupling. 
		Time is 
		measured with respect to $t_{\rm peak}$, the time where $|\Psi_{4,22}|$
		is maximum.
\label{fig:psi4_abs}
}
\end{figure*}

The gravitational quasinormal modes of rotating black holes were 
computed numerically 
in 
Ref.~\cite{Pierini:2022eim} by performing a slow-rotation
expansion, as in Ref.~\cite{Pierini:2021jxd}, to second order in the dimensionless spin
parameter $a\equiv J/m_{\rm BH}^2$ (where $J$ is the angular momentum of black hole)
\footnote{We note that more recently Ref.~\cite{Chung:2024ira} 
developed a general method using
perturbative spectral expansions to compute quasinormal modes in a wide class of 
modified theories of gravity for black holes of any sub-extremal spin
and applied this method to EsGB gravity.}.
According to Ref.~\cite{Pierini:2022eim},
the real frequency of the fundamental $\ell=2,\ m=2$
quasi-normal mode of a black hole in EsGB gravity should decrease 
with coupling and the relative change 
should be $\sim 1\%$ for the largest coupling we consider here.
The correction to the decay rate (imaginary frequency) changes from a positive to a negative
correction for the values of spins and couplings we probe and is expected to be negligible.
We note that Ref.~\cite{Pierini:2022eim} found that the expansion in $\lambda/m_f^2$, 
where $m_f$ is mass of remnant black hole, is accurate within $1\%$ as long as
$\lambda/m_f^2 < 0.07 $ for the real modes and $\lambda/m_f^2 < 0.053 $
for the imaginary modes.  
The couplings we consider correspond to
$\lambda = 0.032m_f^2$ and $0.066m_f^2$, meaning the results here should be applied with
care\footnote{We also note that Ref.~\cite{Pierini:2022eim} 
argued that, although the results are only accurate
to second order in spin, with an appropriate resummation of the spin expansion
parameter, the results
should be accurate for dimensionless spins as large as $\sim 0.7$. The results quoted
here were obtained using the fitting formula Eq.~(47) in Ref.~\cite{Pierini:2022eim}
which do not include the resummation.
Note also that
the computation of the quasinormal modes in 
Refs.~\cite{Pierini:2022eim,Pierini:2021jxd,Blazquez-Salcedo:2016enn} were performed in
Einstein-dilaton-Gauss-Bonnet gravity, which is equivalent to the EsGB
gravity theory considered in this work only in the limit where $\phi$ is small.}.
We find that the real frequency decreases with increasing coupling and 
that the relative 
change is on the order of $\sim 1\%$ for the largest coupling considered, 
i.e. it is the right order of magnitude, but it is too small to reliably quantify with our
current numerical data.
The change in the imaginary part is negligible, also in agreement with
perturbation theory.
The most noticeable effect is a suppression
in the amplitude of ringdown gravitational wave signal with increasing coupling
as shown in the left panel of Fig.~\ref{fig:psi4_abs} (by $\approx 10\%$ for
the largest coupling), which is consistent with an increasing amount of
radiation going into the scalar field with increasing coupling.

\begin{figure*}
	\includegraphics[width=0.99\columnwidth,draft=false]{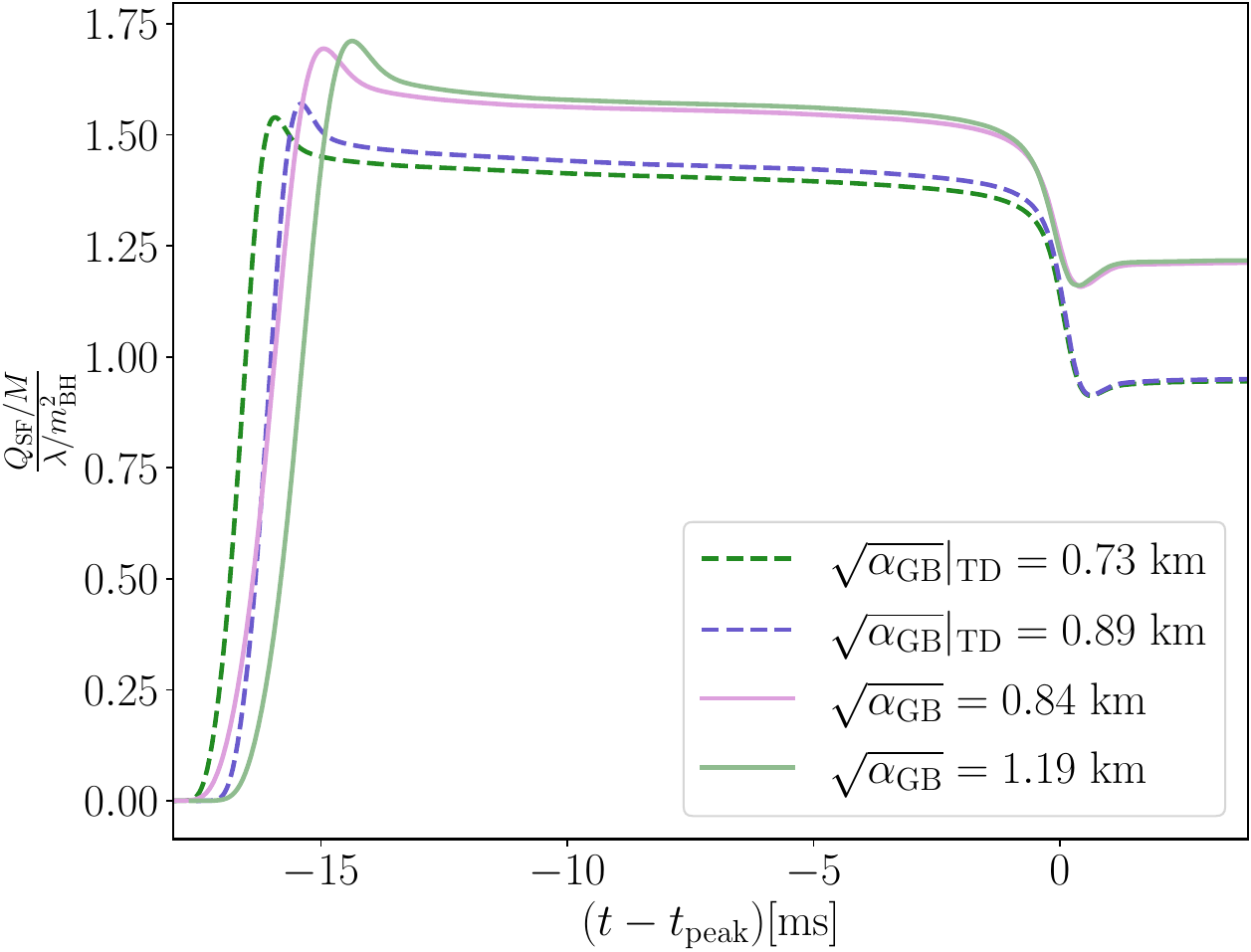}
	\includegraphics[width=0.99\columnwidth,draft=false]{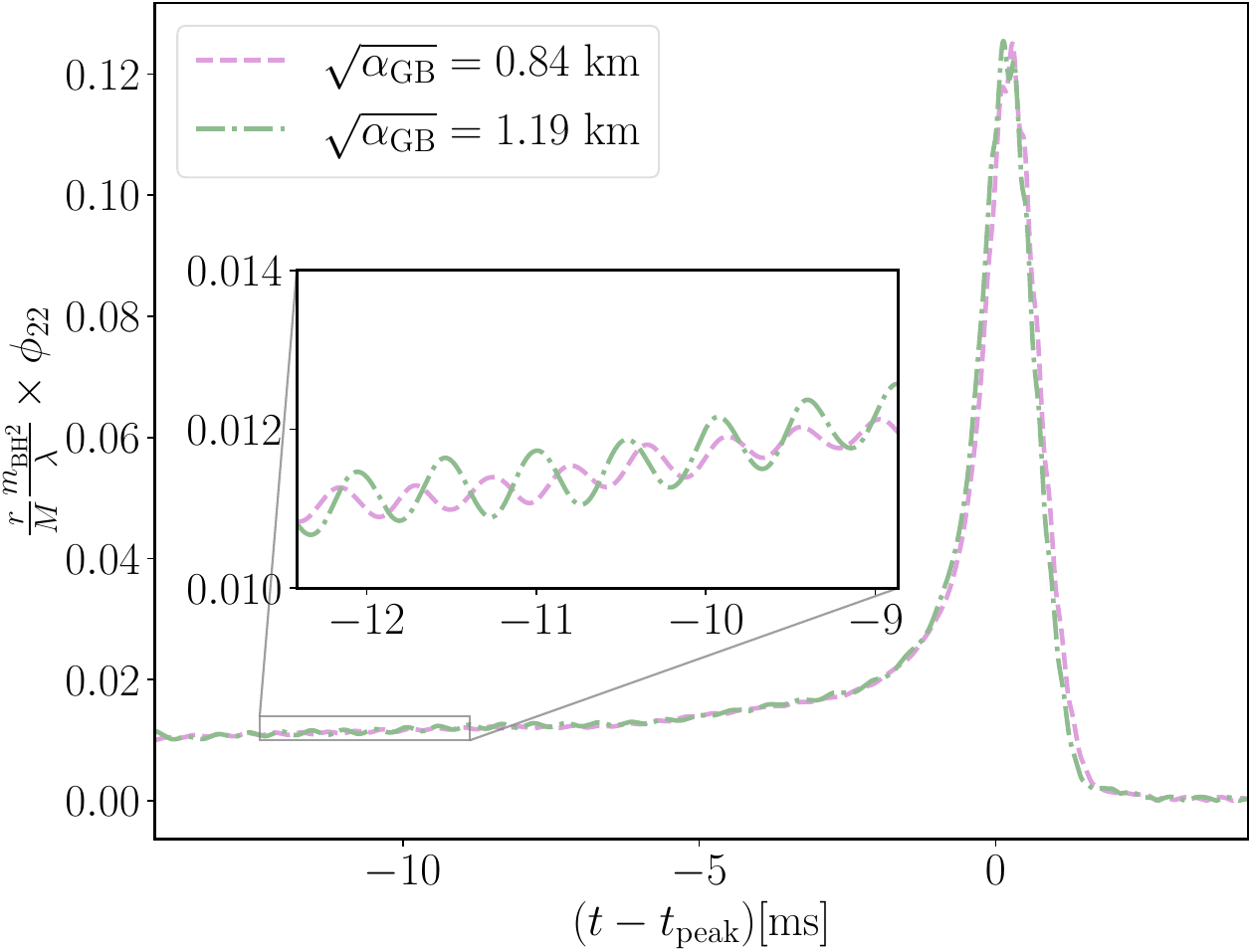}
	\caption{ Left: The scalar charge $Q_{\rm SF}$ 
	rescaled by $\lambda/m_{\rm BH}^2$, measured from the average value of
	the scalar field at $100M$ as a function of retarded time $(t-r)/M$,
	for the BHNS mergers considered in this paper (see 
	the last two rows of Table~\ref{tab:summary}).
	The mergers where the neutron star is tidally disrupted (TD) are indicated
	by dashed lines. 
	Right: Amplitude of
        scalar radiation for a
        BHNS merger with binary parameters consistent
        with the GW200115 event for different values of the EsGB coupling.
        We show the amplitude of the
        $(\ell=2,m=2)$ component of $\phi$ extracted at $100 M$. Time is
        measured with respect to when $|\Psi_{4,22}|$
        is maximum.
\label{fig:GW200225_short_sf_amp}
}
\end{figure*}

\begin{figure*}
	\includegraphics[width=0.99\columnwidth,draft=false]{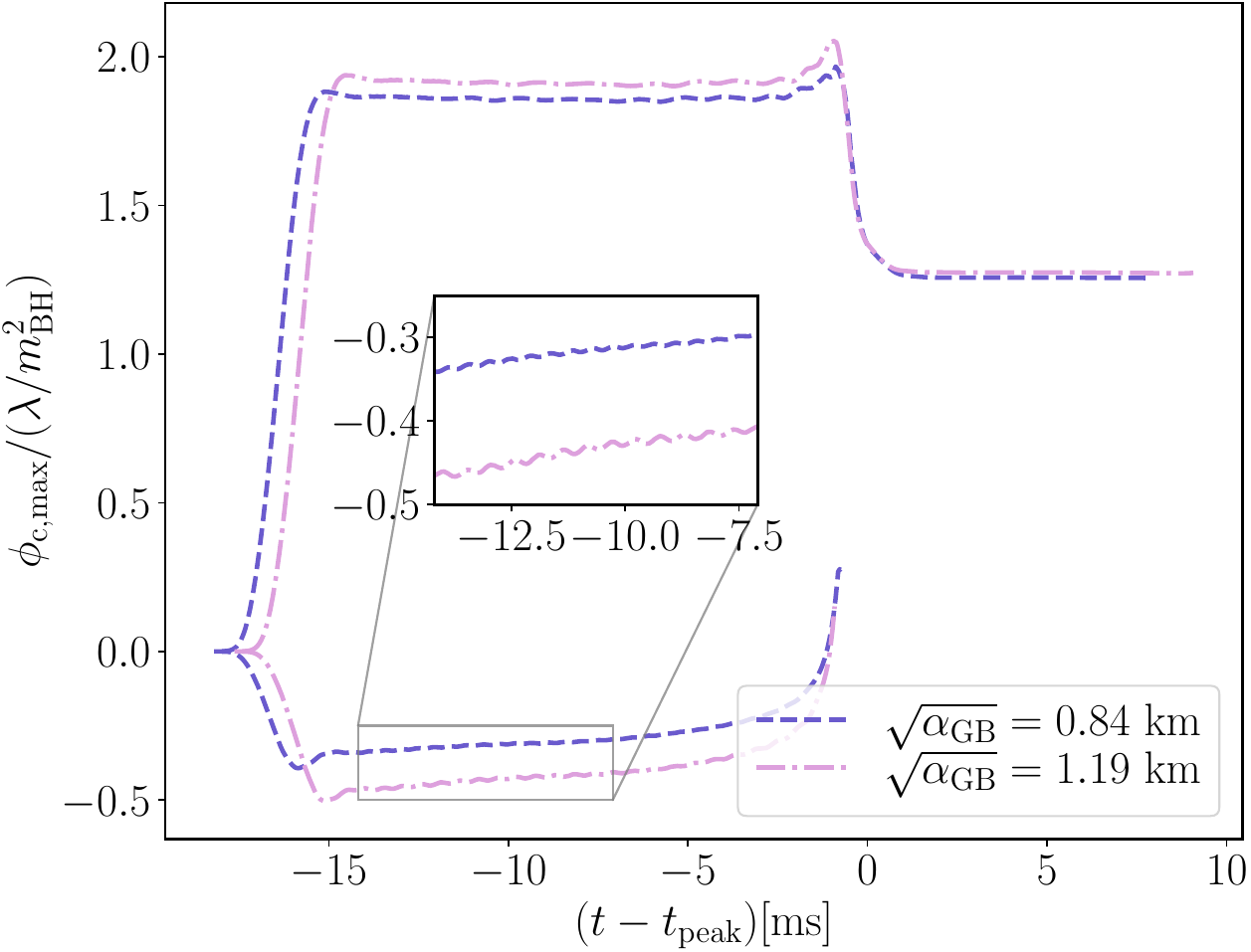}
	\includegraphics[width=0.99\columnwidth,draft=false]{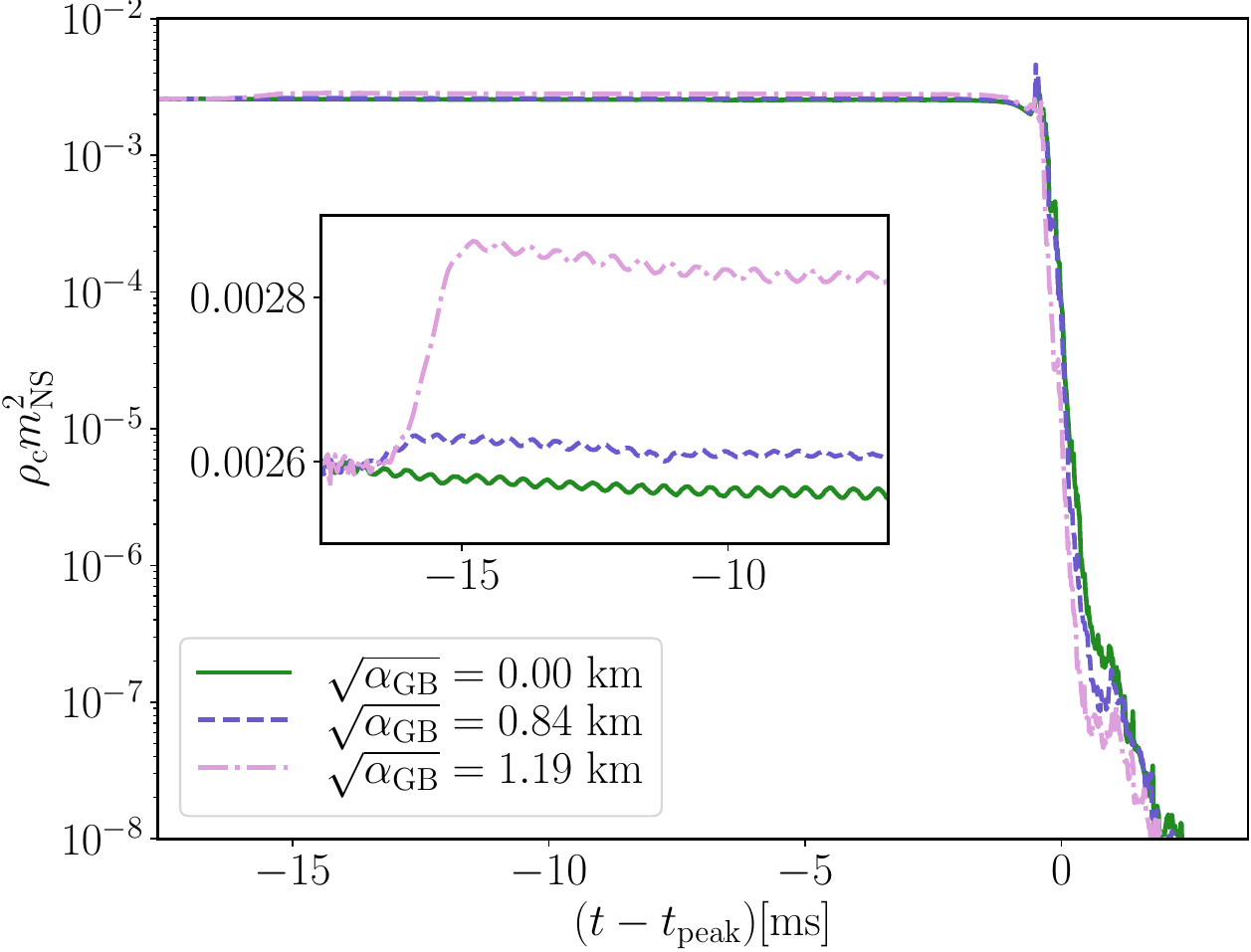}
        \caption{ Left: The maximum value of the scalar field over the
	numerical domain (excluding black hole interior) and value of the scalar field at the center of the neutron
    star (which is negative for most of the evolution). Right: The maximum rest-mass
	density (located at center neutron star).
    These results are for the GW200115-like BHNS merger and various EsGB coupling values.
\label{fig:GW200225_short_phi_rho}
}
\end{figure*}

The amplitude of the $\ell=2,\ m=2$ mode of the scalar waves, shown in 
Fig.~\ref{fig:GW200225_short_sf_amp}, displays small, yet measurable, oscillations
that track
the neutron star oscillations in
the fundamental fluid mode (f-mode) of the star.
These are evident in the oscillations of the star's central value of both the rest mass density
and the scalar field value $\phi_{\rm c}$ (see 
Fig.~\ref{fig:GW200225_short_phi_rho}). 
The relative amplitude of the oscillations in
the central density are not strongly affected by the value of the GB coupling and hence do not
seemed to be an artifact of the way we turn on the scalar field. Instead, we attribute
the oscillations to numerical errors as they decrease as the numerical resolution is increased.
Finally, we note that the right panel of Fig.~\ref{fig:GW200225_short_phi_rho} 
shows that turning on the EsGB coupling increases the central density of neutron star,
up to $10\%$ for the largest coupling we consider,
as predicted from numerical studies of single neutron stars in EsGB \cite{Pani:2011xm}.

\begin{figure*}
	\includegraphics[width=0.99\columnwidth,draft=false]{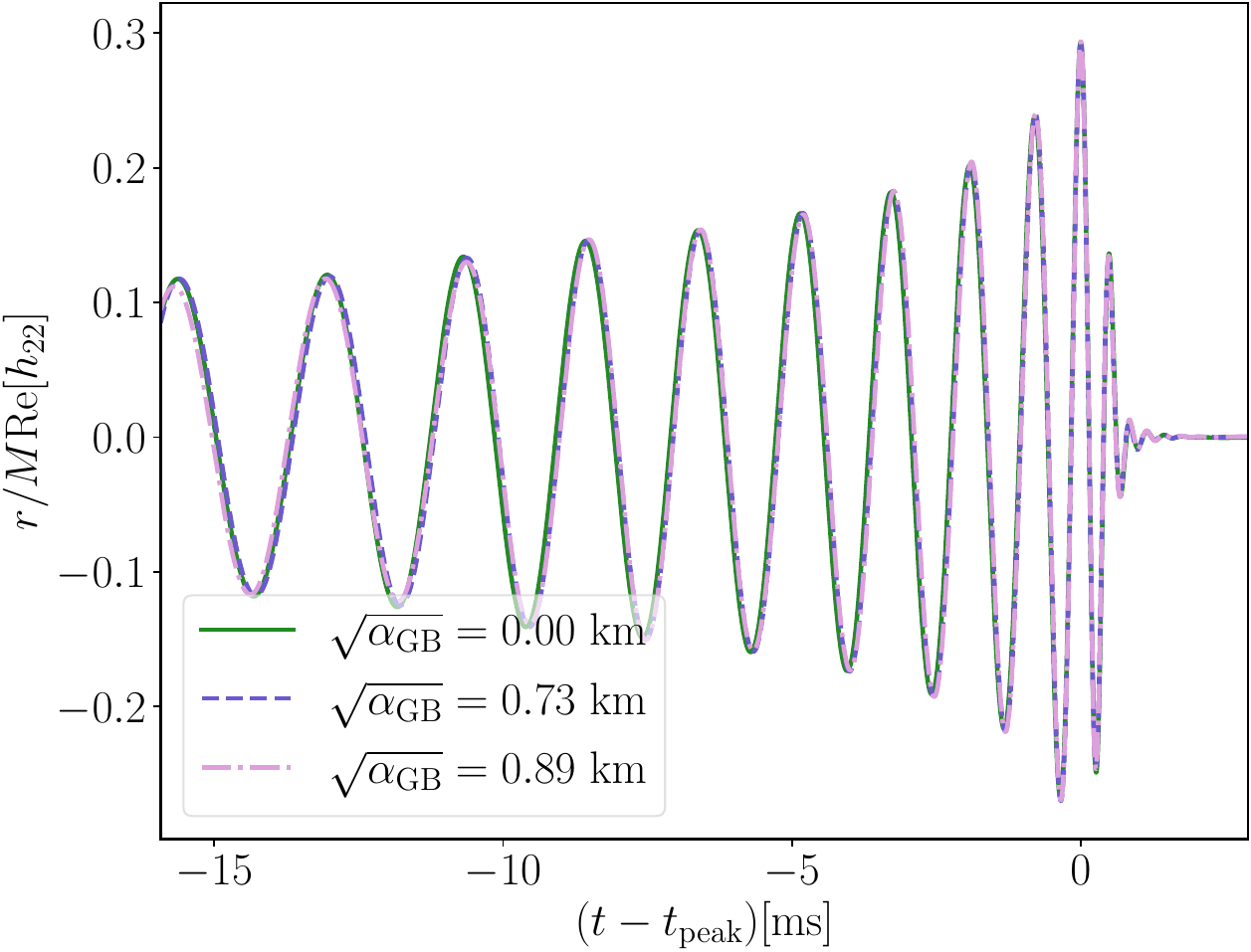}
	\includegraphics[width=0.99\columnwidth,draft=false]{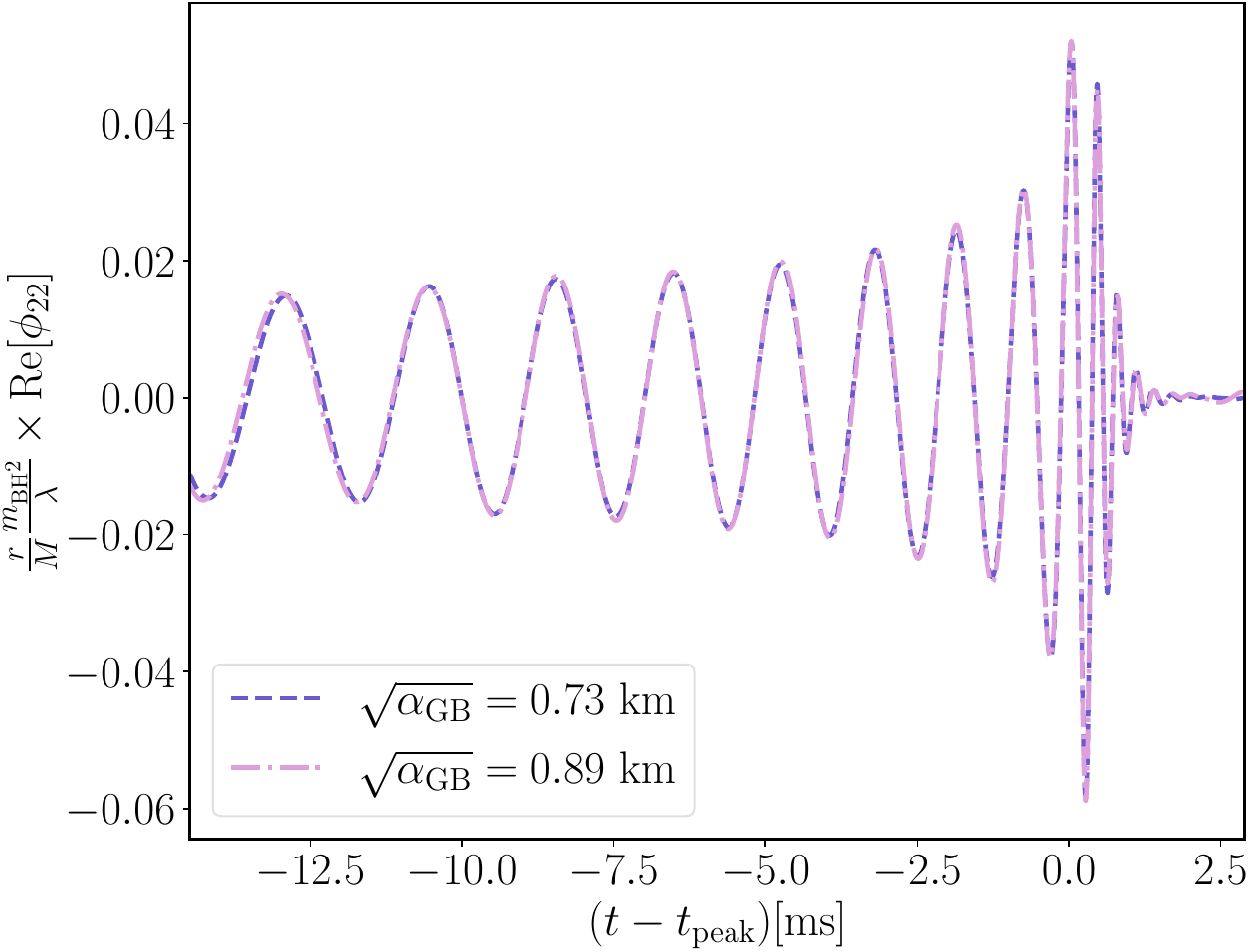}
        \caption{
		Gravitational wave radiation (left) and 
		scalar radiation (right) for a 
		BHNS merger where the neutron star is tidally
		disrupted for different values of the EsGB coupling. 
		We show the real part of
	        the $(\ell,m)=(2,2)$ spherical harmonic of the strain $h$
		and $(2,2)$ component of $\phi$ both extracted at $100 M$. Time is 
		measured with respect to the time where $|h_{22}|$
		is maximum.
\label{fig:q2p5_wave}
}
\end{figure*}

We also consider a BHNS merger where the neutron
is tidally disrupted before merger, with binary parameters similar to the GW230529 event(see last row of Table~\ref{tab:summary}).
We show the gravitational and scalar radiation in Fig.~\ref{fig:q2p5_wave}. The binary
here undergoes $\sim 5$ orbits before merger. We consider evolutions 
with coupling values of $\sqrt{\alpha_{\rm GB}} \sim 0.73$ and $0.89$ km,
where the highest coupling is $50\%$ larger, when normalized by the black hole mass,
than considered in the previous case.
Similarly to before, we find that the most noticeable effect is the decrease in the inspiral
timescale with increasing coupling.
According to perturbation theory, the larger the 
dimensionless coupling value $\lambda/m_f^2$ and spin of remnant black hole, the larger
the change in real and imaginary frequencies of quasinormal modes. 
In comparison to the previous case, the final remnant here is smaller, leading 
to higher coupling values of $\lambda = 0.054m_f^2$ and $0.082m_f^2$, and the final
dimensionless spin is $\sim 0.6$, while it was previously $0.5$. Both of these effects lead to
a larger change in the quasinormal frequency, with a predicted relative change in the 
real frequency of $-2\%$
for the largest coupling we consider and a change of $-0.65\%$ for the decay rate,
according to Ref.~\cite{Pierini:2022eim} (and with the caveats discussed above).
Although we find that the change in the real frequency has the right order of magnitude
and the change in imaginary frequency has the right sign, they are both 
still too small to reliably quantify. 

\begin{figure}
	\includegraphics[width=\columnwidth,draft=false]{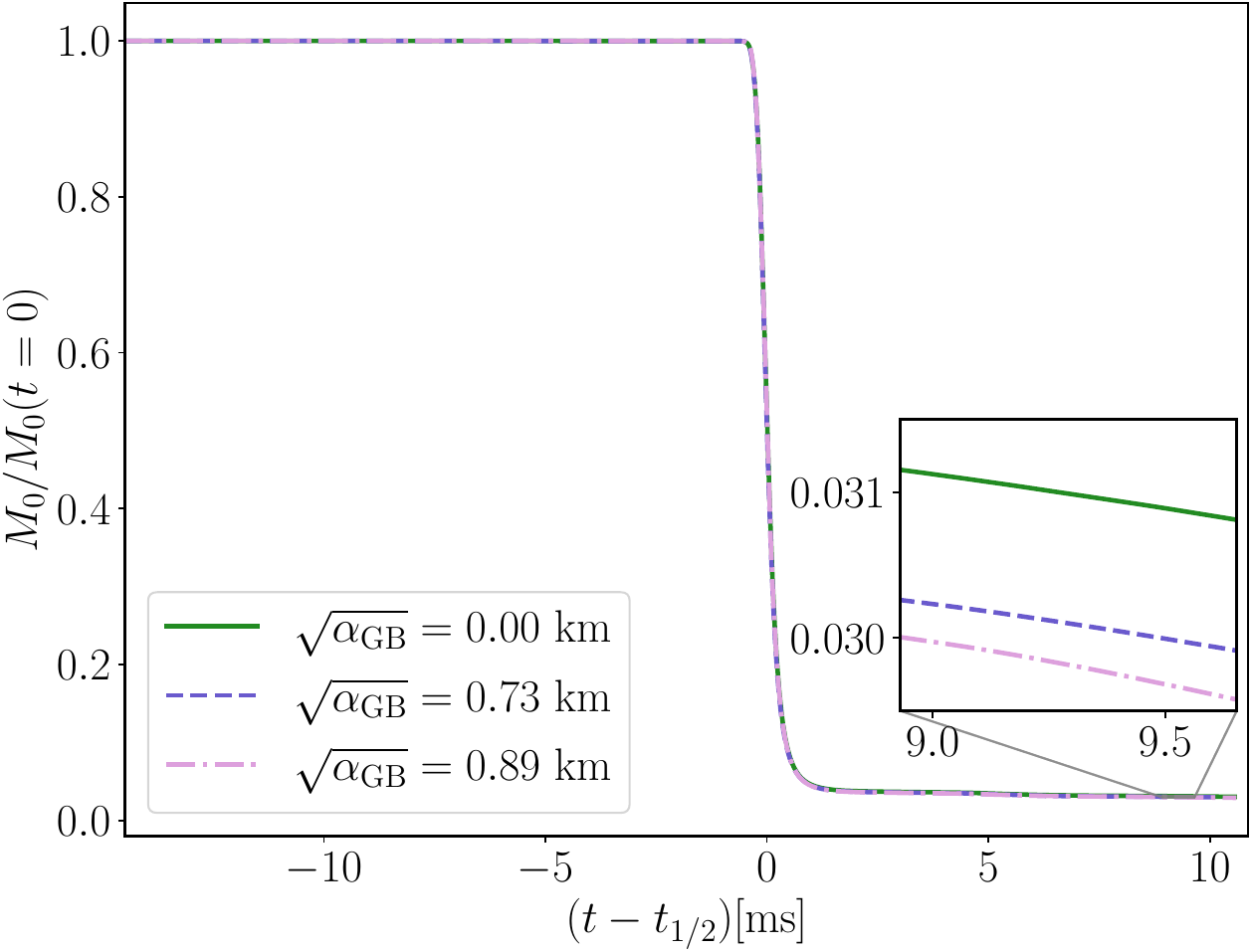}
	\caption{ The total fluid rest mass outside black hole horizon
	for the merger
	where the neutron star is tidally disrupted for different values of the 
	EsGB coupling. Time is measured with respect to the time where half of the neutron
	star mass has been accreted onto black hole $t_{1/2}$ and the rest mass
	is normalized by its initial value.
\label{fig:q2p5_rest_mass}
}
\end{figure}

The main difference compared to the binary where the neutron star is not tidally disrupted 
is that the amplitude of
the ringdown gravitational wave signal increases slightly 
with increasing coupling (by $\approx2\%$ for the largest coupling), as shown in the right panel
of Fig.~\ref{fig:psi4_abs}. 
We hypothesize that this can be attributed to the fact that the neutron star is more compact, 
and less strongly tidally disrupted for larger couplings.
Consistent with this, we also find a small increase in the amount of fluid 
rest mass falling into the black hole with increasing coupling, as shown in 
Fig.~\ref{fig:q2p5_rest_mass}. 
We find that the amount of mass remaining outside of the black hole $8 \ \rm{ms}$
after merger decreases
from a value of $0.049 M_{\odot}$ in GR to $0.047 M_{\odot}$ for the largest coupling
we considered ($3.5\%$ decrease). Of this post-merger material, we estimate that $\sim10^{-3}\ M_{\odot}$
is gravitationally unbound from the system with mildy relativistic asymptotic velocities; we find
no clear trend in the unbound material with EsGB coupling values.
In passing, we note that the leftover rest mass in GR is roughly an order of magnitude
larger than that predicted by the fitting formula in Ref.~\cite{Foucart:2018rjc}.
Though some of the discrepancy may be due to truncation error, the formula in Ref.~\cite{Foucart:2018rjc}
was also not fit with any simulation results in the range $q^{-1}\in(1.2,3)$, making
it difficult to judge its uncertainty in this regime.  

%=============================================================================

%=============================================================================
\section{Discussion and Conclusion}
\label{sec:conclusion}
In this work, we have taken advantage of recent advances in solving the
full
equations of EsGB gravity to study BHNS mergers for the
first time. This was motivated by the fact that neutron stars do not
have scalar charge in this theory and black hole masses
in such binaries are typically small in comparison to binary black
holes \cite{Biscoveanu:2022iue}, making them an ideal probe to test
for modifications to GR at smaller curvatures length scales.
We find that the BHNS binaries inspiral faster in EsGB
relative to GR due to the emission of scalar radiation.
We first evolved a system chosen to be consistent with GW200115 and with
EsGB coupling at the limit of the observational bounds placed by applying PN predictions to the
event. Comparing our scalar and gravitational waveforms to existing
PN predictions for EsGB, we find reasonable agreement in the dephasing
relative to GR all the way up to end of $\sim 500$ Hz,
in part due to the fact that the dephasing between GR and EsGB becomes small in the final
orbits. This suggests current
bounds on EsGB using PN theory up to the end of inspiral phase are a good
approximation.
It would be interesting to carry out a full Bayesian parameter estimation
using PN theory as well as the theory agnostics approaches such as
TIGER or FTI to better understand how these methods perform and study
some of the degeneracies that might arise and lead to parameter estimation
biases. To fully understand the observational prospects of constraining EsGB gravity
with BHNS mergers, future work should also explore a range of
EOSs, EsGB couplings, and binary parameters, and understand possible
degeneracies, including with tidal effects~\cite{Ma:2023sok}. 

In addition to measuring the dephasing, we also found that the leading order PN
contribution compares well in matching the amplitude of scalar radiation 
emitted during the inspiral at a given frequency. This is in 
qualitative agreement with binary black hole simulations in EsGB \cite{Corman:2022xqg},
and can be partially explained by the fact that corrections to the scalar field
amplitude in the GB coupling enter at third order for shift-symmetric EsGB
gravity (see, e.g., Appendix D of \cite{Corman:2022xqg}). We also
note that the next-to-leading order was found to increase the error in amplitude,
and the next-to-next-to-leading order is needed to improve consistency.

We also studied the effect of modifications to GR
on the dynamics of the merger and ringdown signal of newly formed black hole for
two different scenarios: a BHNS merger where the neutron star
is not tidally disrupted, and one where it is. Most of the literature has focused
on computing a change in frequency using perturbation theory. However,
for both cases considered here, we find that the frequency shift
is small, in qualitative agreement with perturbative predictions, and we find that 
the dominant effect is instead a change in the amplitude of the ringdown signal. 
This observation is in agreement with evolutions of binary black hole and
binary neutron star mergers in EsGB \cite{Corman:2022xqg,East:2022rqi,
Evstafyeva:2022rve}.
This observational signature could potentially be leveraged in ringdown tests of GR, 
but also introduces further complications \cite{Maggio:2022hre,Gennari:2023gmx,
Maselli:2019mjd,Ghosh:2021mrv,Silva:2022srr,Carullo:2021dui,Maselli:2019mjd}. 
In particular, we found a suppression of the amplitude with increasing coupling
when the neutron star is not tidally disrupted, explained by an increase by
the amount of emitted scalar radiation. However, when the neutron star is tidally
disrupted, we found that the amplitude increases with coupling, which we attribute
to the fact that, for a fixed EOS, the NS is more compact and less easily tidally-disrupted.

Even setting aside modified gravity considerations, the lower mass ratio case we consider ($q=0.4$) 
probes a regime that has not been extensively studied using full GR simulations of BHNS mergers,
but has become particularly interesting with the observation of GW230529. It is worth
noting that we find (in GR) that an accretion disk forms post-merger with a few percent of a solar mass, which is one order
of magnitude larger than the prediction of the commonly used 
fitting formula in Ref.~\cite{Foucart:2018rjc} for this particular case. 
Coupled with other recent studies~\cite{Chen:2024ogz,Martineau:2024zur}, this suggests
that analyses using this formula may underestimate the prospects for a post-merger electromagnetic transient
in this part of the parameter space. 

Though not comprehensively addressed here, it would also be interesting for
future work to quantify how modified gravity affects potential electromagnetic
transient arising from the merger. This would involve considering a range of
binary parameters and choices for the neutron star EOS in order to determine
under what circumstances modified gravity effects could be important, and
non-degenerate with other parameters, in determining the size of the post-merger
accretion disk, the amount of unbound material, and other properties that
affect potential electromagnetic signatures.

%=============================================================================
\section*{Acknowledgements}
We thank Erik Schnetter and Samuel Toothe for their assistance constructing 
the initial data used here. M.C. is grateful to Harald Pfeiffer and Elise S\"anger for 
helpful discussions regarding various aspects of this project.
M.C. is especially thankful to
F\'elix-Louis Juli\'e for sharing his notebook on how to transform the
scalar field from the Jordan to the Einstein frame as well as answering various
questions regarding PN calculations in EsGB gravity. We thank Michalis Agathos for
reviewing this manuscript.
W.E. acknowledges support from a Natural Sciences and Engineering Research 
Council of Canada Discovery Grant and
an Ontario Ministry of Colleges and Universities Early Researcher Award.
This research was
supported in part by Perimeter Institute for Theoretical Physics. Research at
Perimeter Institute is supported in part by the Government of Canada through
the Department of Innovation, Science and Economic Development and by the
Province of Ontario through the Ministry of Colleges and Universities. 
This research was enabled in part by support
provided by SciNet (www.scinethpc.ca) and the
Digital Research Alliance of Canada (alliancecan.ca). Calculations were
performed on the Symmetry cluster at Perimeter Institute, 
the Niagara cluster at the University of 
Toronto, and the Narval cluster at Ecole de technologie sup\'erieure 
in Montreal.
Computations were also performed on the
Urania HPC system at the Max Planck Computing and Data Facility.
This material is based upon work supported by NSF's LIGO Laboratory which is a major 
facility fully funded by the National Science Foundation.

\newpage
%=============================================================================
\appendix
%=============================================================================
\section{\label{app:convergence} Numerical convergence and error estimates
}
\begin{figure*}
	\includegraphics[width=0.99\columnwidth,draft=false]{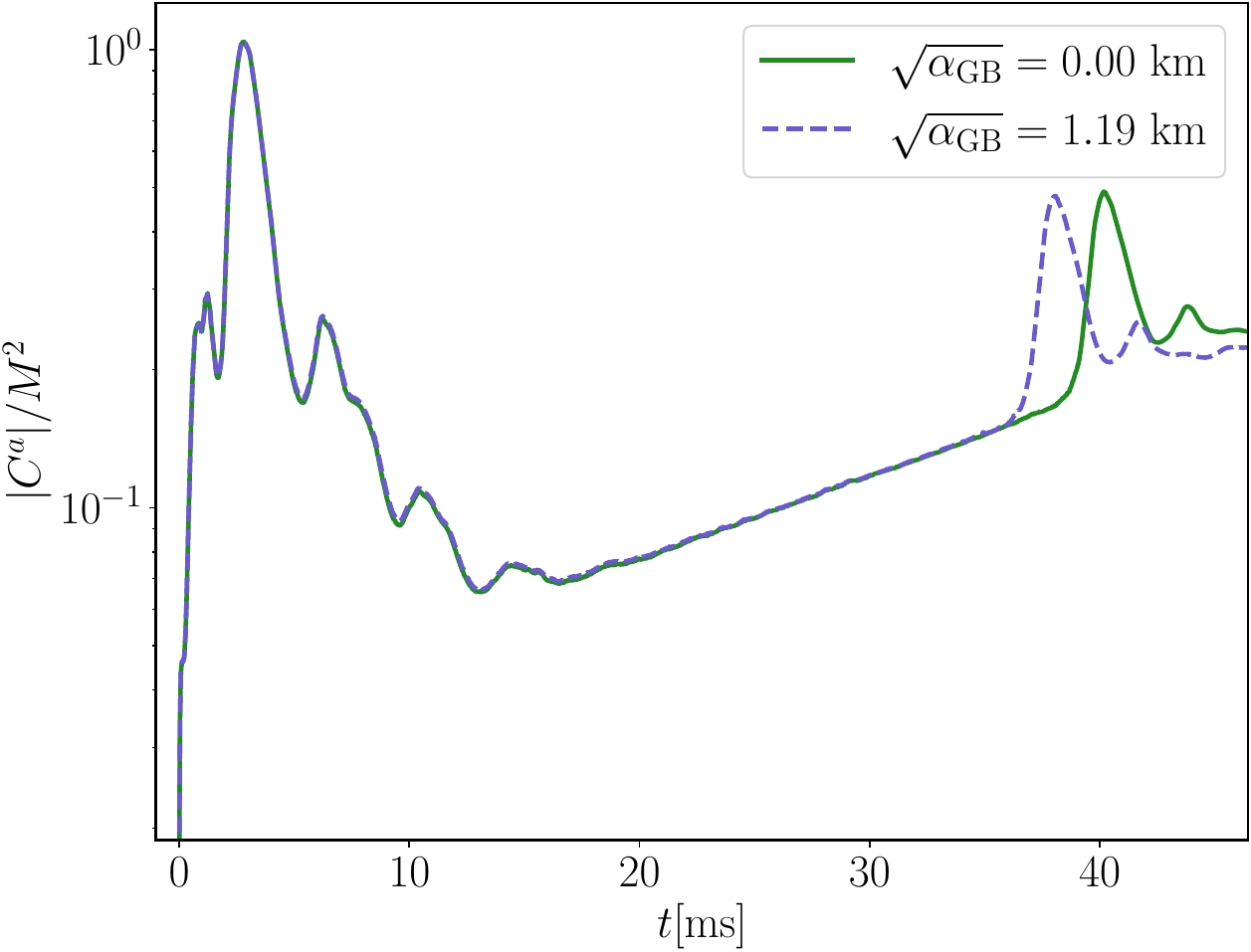}
	\includegraphics[width=0.99\columnwidth,draft=false]{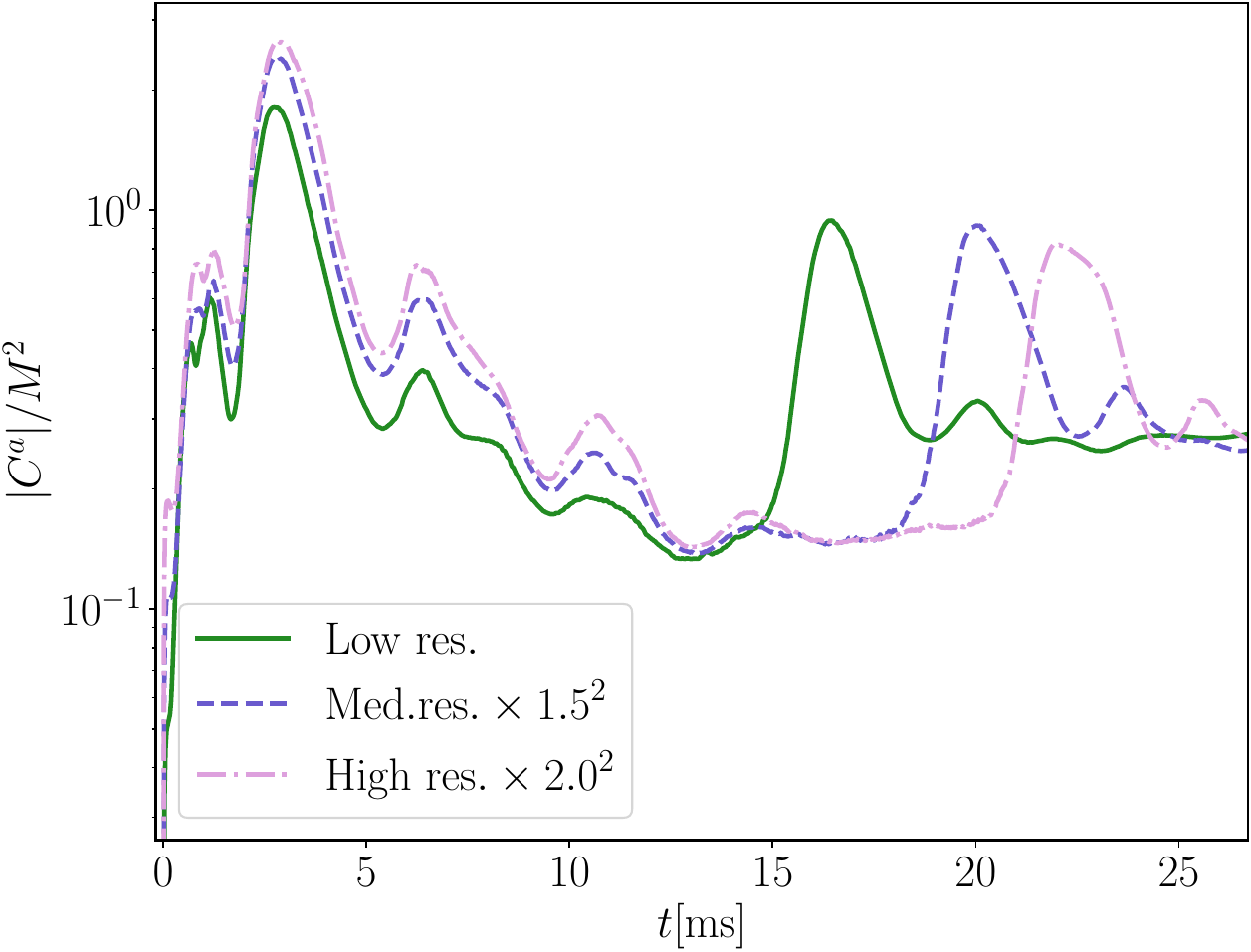}
	\caption{Left: Volume integrated norm of the 
	modified generalized harmonic constraint violation $C^a$ as a function
	of time for the BHNS system with GW200115-like parameters
	and an initial separation of $D=9.8 M$ in GR and EsGB with 
	$\lambda=0.1 m_{\rm BH}^2$.
	We observe that the constraints are the same modulo a time shift. Right: Convergence
	of the volume integrated constraint violation for same system as in left panel,
	but for an initial separation of $D=8.6M$ and at three resolutions. The values
	have been scaled assuming second order convergence, though at early times the
	convergence is closer to first order.
\label{fig:cnst_long}
}
\end{figure*}

For the BHNS mergers considered in this paper, we perform
simulations with seven levels of refinement where the finest level has a linear grid spacing
of $dx \sim 0.016 M$, 
and each successive level has a linear grid spacing that is twice as coarse.
In Fig.~\ref{fig:cnst_long}, we show the norm of the modified generalized harmonic 
constraint violation, integrated over the domain as a function of time.
In the left panel, we compare a simulation that transitions to a nonzero EsGB coupling to the equivalent simulation in GR.
As can be seen, this transition does not noticeably impact the constraint violation, modulo
the faster inspiral of the modified gravity system.  

For the GW200115-like binary where the initial separation is $D=8.6 M$, we also 
perform a convergence study with grid spacing that is $4/3$ and $2/3 \times$ as large,
which is shown in right panel of Fig.~\ref{fig:cnst_long}.
All results in the main text are from the medium resolution.
Although at early times the order of convergence is closer to first order,
presumably from high frequency noise (junk radiation) in the initial data
which may engage the shock capturing scheme,
at later times the convergence is consistent
with roughly second order, as expected from our numerical scheme in the absence of shocks.
In addition, we note that the constraints jump again at the end, which
corresponds to when the NS starts to plunge into the
BH.

We compare the dephasing between the EsGB and GR waveforms in 
Fig.~\ref{fig:GW200115_long_gw} 
to the numerical errors in the simulations using the
techniques applied to binary black hole mergers in Ref.~\cite{Corman:2022xqg} 
(and detailed
in Appendix A of that paper). The error in the Richardson extrapolated phase 
at the frequency where GR peaks is $\sim 7.2$ radians, 
which is comparable to the dephasing.
However, similarly to the binary black hole mergers in Ref.~\cite{Corman:2022xqg}, we find that
the dominant truncation error in our simulations does not depend strongly on the value
of the coupling and therefore partially cancels out when calculating the difference
in gravitational wave phase between EsGB and GR simulations using the same resolution.
We see evidence that this is the case, 
for example, by comparing a measure of the truncation
error in $\Delta \Phi$, computed by comparing the GW200115 simulation starting at a 
shorter 
initial separation in GR to an equivalent EsGB simulation with 
$\sqrt{\alpha_{\rm GB}} = 1.19\ \rm{km}$ at two different resolutions,
to an estimate of the overall truncation
error in $\Phi$ for the same EsGB case. 
We find the former to be $\sim 6 \times$ smaller than the latter 
(see Fig.~\ref{fig:error_cancellation} in Appendix~\ref{app:convergence}). 

\begin{figure*}
	\includegraphics[width=0.99\columnwidth,draft=false]{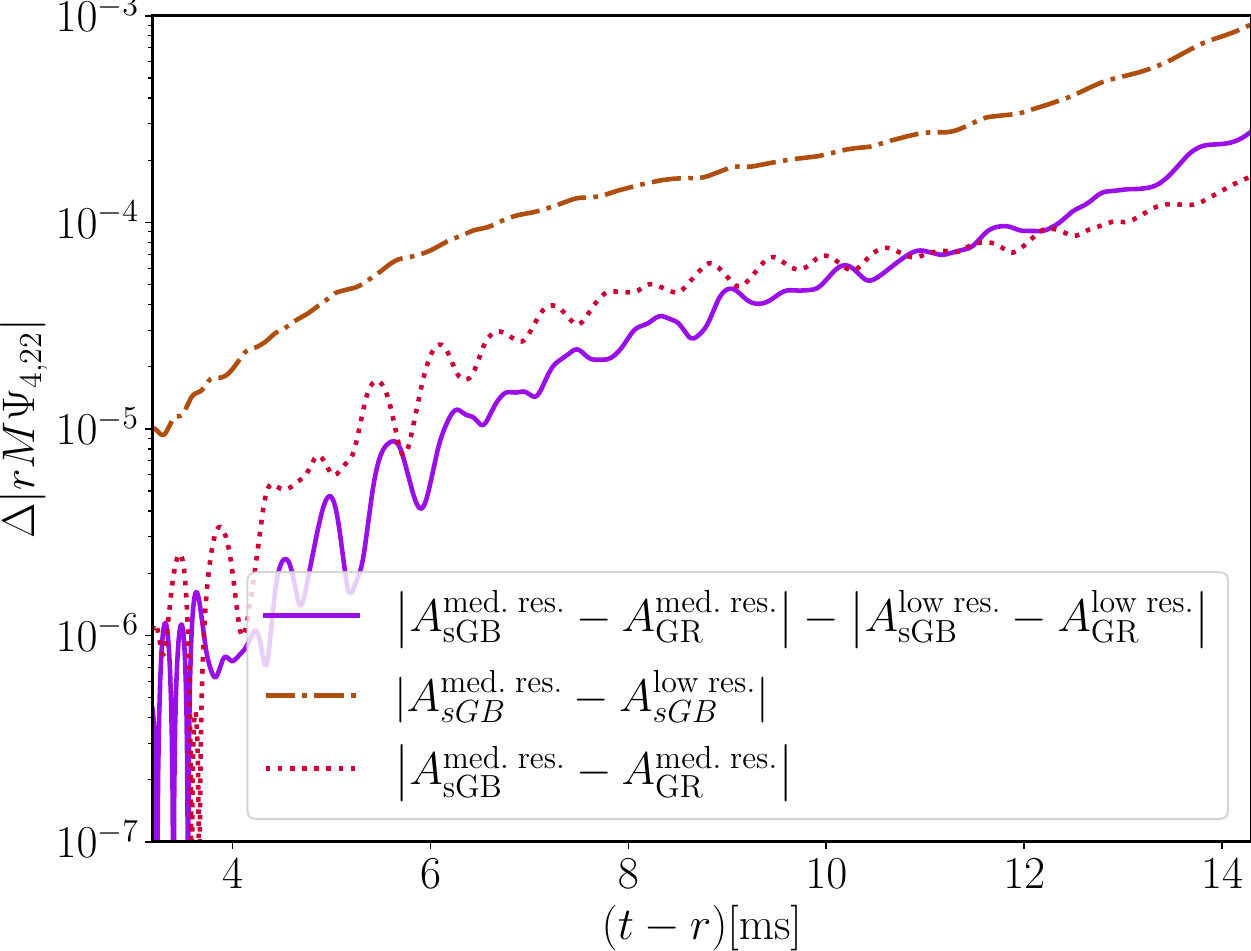}
	\includegraphics[width=0.99\columnwidth,draft=false]{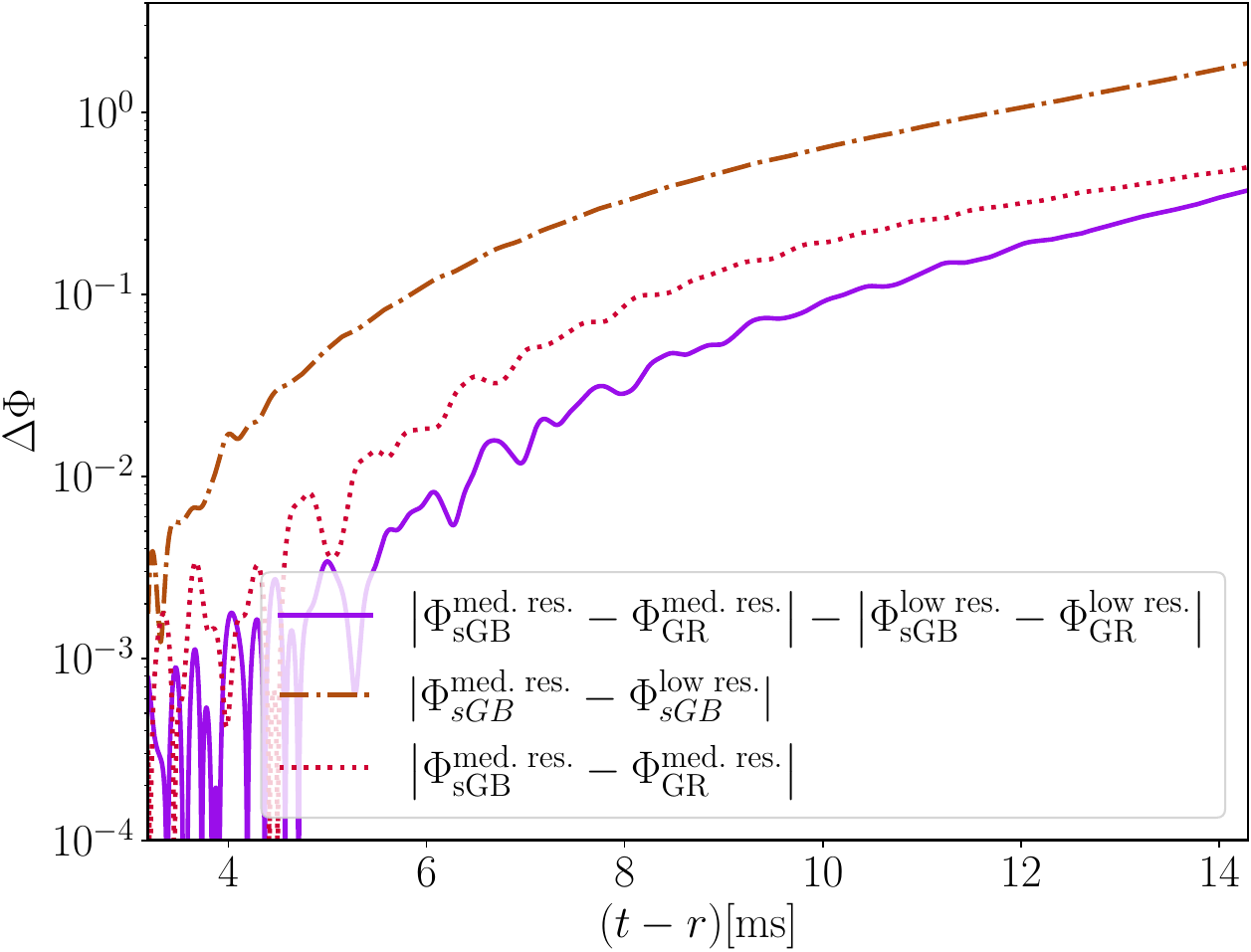}
        \caption{
		We show the difference between the low and medium resolutions 
		of the amplitude (left) and phase (right) of the gravitational waveform 
		for the BHNS binary with GW200115 like parameters, 
		an initial separation of $D=8.61M$, and coupling of 
		$\sqrt{\alpha_{\rm GB}} = 1.19 \rm{km}$.
		We also show the difference between the
		EsGB and GR amplitude and phase at low and medium resolutions 
		(dashed brown line). This
		provides evidence that the truncation error roughly cancels between the 
		EsGB and GR runs.
\label{fig:error_cancellation}
}
\end{figure*}

%=============================================================================
\bibliography{./main}

\end{document}